\documentclass[range]{ar2e}
\usepackage{ARAstroBib}
\usepackage{graphicx}

\newcommand{\teff}{$T_{\rm eff}$}
\newcommand{\logteff}{\mbox{$\log T_{\rm eff}$}}
\newcommand{\vsini}{\mbox{$v \sin i$}}
\newcommand{\prot}{$P_{\rm rot}$}
\newcommand{\bv}{$(B-V)$}
\newcommand{\bvz}{$(B-V)_0$}
\newcommand{\vi}{$(V-I)$}

\newcommand{\Lbol}{\mbox{$L_{\rm bol}$}}
\newcommand{\loglbol}{\mbox{$\log L_{\rm bol}$}}
\newcommand{\MS}{$M_\odot$}

\newcommand{\apj}{{\it Astrophys. J.}}
\newcommand{\apjs}{{\it Astrophys. J. Suppl. Ser.}}
\newcommand{\aj}{{\it Astron. J.}}
\newcommand{\aap}{{\it Astron. Astrophys.}}
\newcommand{\aaps}{{\it Astron. Astrophys. Suppl. Ser.}}
\newcommand{\aaprev}{{\it Astron. Astrophys. Rev.}}
\newcommand{\araa}{{\it Ann. Rev. Astron. Astrophys.}}
\newcommand{\mnras}{{\it Mon. Notices Roy. Astron. Soc.}}
\newcommand{\memital}{{\it Mem. Soc. Astr. Ital.}}
\newcommand{\annphys}{{\it Ann. Phys.}}
\newcommand{\pasp}{{\it Pub. Astr. Soc. Pac.}}
\newcommand{\physscr}{{\it Phys. Scr.}}
\newcommand{\solarphys}{{\it Solar. Phys.}}
\newcommand{\apss}{{\it Astrophys. \& Space Sci.}}
\newcommand{\nature}{{\it Nature}}
\newcommand{\science}{{\it Science}}
\newcommand{\revmodphys}{{\it Rev. Mod. Phys.}}

\begin{document}

\input epsf.tex    
\input epsf.def   

\input psfig.sty

\jname{ARAA}
\jyear{2010}
\jvol{TBD}
\ARinfo{TBD}

\title{The Ages of Stars}

\markboth{The Ages of Stars}{David R. Soderblom}

\author{David R. Soderblom
\affiliation{Space Telescope Science Institute}}

{\bf The following paper will appear in the 2010 volume of {\it Annual Reviews of Astronomy and Astrophysics.}}


\begin{abstract}
The age of an individual star cannot be measured, only estimated through mostly model-dependent or empirical methods, and no single method works well for a broad range of stellar types or for a full range in age.  This review presents a summary of the available techniques for age-dating stars and ensembles of stars, their realms of applicability, and their strengths and weaknesses.  My emphasis is on low-mass stars because they are present from all epochs of star formation in the Galaxy and because they present both special opportunities and problems.  The ages of open clusters are important for understanding the limitations of stellar models and for calibrating empirical age indicators.  For individual stars, a hierarchy of quality for the available age-dating methods is described.  Although our present ability to determine the ages of even the nearest stars is mediocre, the next few years hold great promise as asteroseismology probes beyond stellar surfaces and starts to provide precise interior properties of stars and as models continue to improve when stressed by better observations.

\end{abstract}

\maketitle


\section{WHY AGES MATTER}

We can directly measure many key physical properties of a star.  The masses of stars in binary orbits can be determined from basic physics, and we can infer the mass of most single stars to $\sim10\%$ just from a spectral type, particularly for the main sequence.  Determining the composition of a star is not easy, but it is straightforward, and the process and its limitations are well understood.  But age, the third key determinant of a star's physical state, is another matter all together.  The Vogt-Russell theorem tells us that the physical state of a star results from its mass and composition.  Other factors (rotation, companionship, magnetic fields) matter too, but mass and composition dominate.  The composition of a star changes with time, and so age influences the state of a star, but less directly.  Time is a medium in which the star inexorably changes, but age is not the direct agent of that change.

All of this means that a star's age cannot be measured, it can only be estimated or inferred.  This is underscored by the fact that we have exactly one stellar age that is both precise and accurate, and that, of course, is for the Sun.  But the Sun itself does not reveal its age; it is only because we can study solar system material in the laboratory that we can limit the Sun's age.  We can do that for no other star.

\marginpar{\bf Sidebar 1: Why Care About Stellar Ages?\\ (see p. \pageref{sidebar1})}

Yet age lies at the heart of stellar evolution, for how can there be evolution without the passage of time?  Sometimes our desire to know ages is satisfied if we can place events in the correct sequence and understand the relative duration of phases.  But often we need to tie together observations of independent objects to arrive at an understanding of a process.  Some examples include:

\begin{itemize}

\item The formation and evolution of proto-planetary disks appear to occur in the first $\sim100$ Myr of a star's life, with debris disks forming later.  At present we can just barely limit this time-scale with the methods available, but clearly one would like to be able to see differences in the formation processes and time-scales from place to place and star to star if we are to understand fully the physics of the processes involved.

\item Many gas-giant planets have now been found around nearby solar-type stars.  We can estimate rough ages for those stars, but a better understanding of, for instance, the dynamics of such systems and the inward migration of planets requires accurate ages.

\item We could better understand the behavior and future of our Sun if we could create a true cohort based on mass, composition, and age.  At present, comparisons of the Sun to other stars often rely on the very stellar properties (such as activity, rotation, or lithium abundance) that one is trying to understand; the process is self-referential.

\item To fully understand the Galaxy's history of formation, enrichment, and dynamics, we need to be able to assign ages to individual stars in its components.  Low-mass stars live long enough to be present from all epochs of star formation, making them attractive for these purposes.  Recent studies have applied some of the techniques discussed in this review to Galactic problems, with differing and sometimes conflicting results.

\item The search for life and any understanding gained from observing signs of life beyond our solar system has a profound need to measure stellar ages if we hope to gain insights into biological evolution.

\end{itemize}

These examples come to mind because of my own interests, and the reader can supply many more.  A stellar chronologist will not lack for work.  A full understanding of many phenomena in astrophysics means at least getting the sequence of events right, and, preferably, knowing time-scales.  Many problems are connected to individual stars or the nearest stars, and we need reliable ways to get at the ages of such objects.  I also note that the overall cosmic age scale, which is most unambiguously constrained by the age of the solar system, is perhaps the single most contentious point in all of science for those who are science deniers, and an understanding of that datum and of ages in general is important for us as involved citizens.

This review will discuss a broad range of stars, both as individuals, and as clusters, associations, and groups.  Cluster ages are important in themselves, of course, but here I use clusters mostly as benchmarks to calibrate other age estimators.  I will emphasize late-type stars because they are numerous, astrophysically interesting in themselves, and are present from all epochs of star formation in the Galaxy.

\marginpar{\bf Sidebar 2: A Challenge for Stellar Astrophysics\\ (see p. \pageref{sidebar2})}

\subsection{What Do We Mean By Age?}

When considered in detail, the age of a star is inherently uncertain if only because the point at which its age begins is ill-defined.  Theory might use the point when hydrostatic equilibrium is established, but that is not easy to pinpoint observationally.  One baseline that is reasonably well-defined in the observations of pre-main sequence stars is the deuterium-burning ``birthline'' of \citet{stahler88}, at the point where proto-stars lose their shrouds and reveal themselves.  At that point a 1.0 \MS\ star is $\sim10^5$ years old.  \citet{wuchterl03} have proposed a time zero-point for when an object's photosphere first exists, which they take to be when the Rosseland mean opacity is 2/3.  The \citet{wuchterl03} zero-point precedes the birthline by perhaps $\sim10^4$ years and is dependent on the star's accretion history \citep{froebrich06}.  These differences are critical for fully understanding the earliest phases of star and planet formation, but that degree of uncertainty becomes essentially meaningless for the stars to be discussed here, which are $\sim1$ Myr old and more.

Some have suggested using the arrival of a star on the zero-age main sequence to define ``$t=0$.''  That is a well-defined point in a star's life, both theoretically and observationally, but it is midway through the star's life and would require the use of negative ages.  We cannot yet define well the true birth of a star in a dense cloud, but the birthline is reasonably well-defined, both in models and observationally.  Starting with $t=0$ at that point makes it convenient to work on a logarithmic scale.  As we will see, most of the various age estimators are also best used logarithmically.

The concept of ``age'' is different for white dwarfs (WDs) because the age is based on the cooling of the WD (see Sec. \ref{WDs}).  This calculation is based on well-understood physics, which is what makes WD cooling ages attractive, but the cooling does not start until the WD forms, and the time elapsed up to that point depends on more conventional stellar modeling that is subject to a number of uncertainties.   This is not important for older clusters, where the WD progenitors are massive stars for which the pre-WD phase is a small fraction of the cluster's age.

\subsection{Goals of This Review}

The need for reliable stellar ages has always existed, but recent work has added significantly to the observations available to address these problems, and the quality of the models has kept pace.  The problems are not solved, but we probably have a better understanding of the limitations.  Also, as I note at the end, I expect major advances in this area over the next few years.

I cannot review the full history of this subject, but a key motivating paper was that of
\citet{skumanich72},
who looked at some age-related observable properties of solar-type stars.
He had little data to work with at the time, but the power-law relations he plotted for rotation and activity (and an exponential for depletion of lithium) instigated much subsequent work.

Despite the unambiguous importance of knowing the ages of stars, the subject has been the focus of only a very few topical symposia and few or no previous general reviews.  I will start with an outline of the available techniques that can be applied to individual stars and a scheme for understanding them.    I will emphasize ``Population I'' or ``thin disk'' stars and low- to intermediate mass stars simply because they predominate in the solar neighborhood and are central to many of the questions for which an age is part of the answer.  I hope to improve on the status quo by compiling the available information in one place and by attempting to systematically lay out alternatives and their advantages and shortcomings.  I will compare the different methods when possible, and I will point out areas that can benefit from attention.

This review will focus mostly on late-type dwarfs and pre-main sequence stars.  Elderly individual stars will be discussed, but the roader topic of Population II stars is beyond the scope of this.  Similarly, white dwarfs (WDs) are also very useful for age-dating clusters and populations using fundamentally different physics, but they have been well reviewed elsewhere (\citet{koester02}, \citet{vonhippel05}, \citet{winget08}, \citet{salaris09}) and so will only be mentioned here (see also Sec. \ref{WDs}).  Evolved stars present some special problems of their own and also will not be discussed here.  However, more on the ages of these objects (and on stellar ages in general) can be found in \citet{mamajek09}, which contains review papers from IAU Symposium 258, ``The Ages of Stars.''   Evolved stars, because of their high luminosities, are often used as tracers of populations, and a discussion of the ages of populations can be found in \citet{tolstoy09}.  Non-stellar objects will not be treated here.

I will also briefly re-examine open clusters and cluster ages as the benchmarks of other methods.  The correct age scale to be used for OCs remains fundamentally uncertain to a significant degree, as we will see, but for the most part other age estimators are at least satisfactory if they provide results on a scale that is consistent with the OCs.  Other ways to calibrate age indicators are also discussed.

\subsection{Some Helpful Terms and Abbreviations}

A few abbreviations are used here so often that this is a good place to define them.  ZAMS is, of course, the zero-age main sequence, the point in a star's life when hydrogen starts to fuse into helium in the core.  Pre-main sequence (PMS) stars are young objects that have not yet reached the ZAMS, and main sequence (MS) stars are more generically ZAMS and later.  We distinguish between the Hertzsprung-Russell diagram (HRD), which is in physical units of luminosity and temperature, and the color-magnitude diagram (CMD), which is in observational units.  One is better suited to some discussions than the other, and the transformations between the two are not trivial.  A frequently-discussed locus in either the HRD or CMD is the main-sequence turnoff (MSTO), the region in an OC's HRD where stars stop getting hotter and bluer and an isochrone bends back to the red for more massive stars.  We will use $\tau$ and $\log \tau$ to denote an age, in years.

Open clusters (OCs) are discussed because they are so important as age benchmarks.  OCs are well-defined groups of stars that are clearly in close proximity to one another and which are defined observationally by having only a small dispersion in radial velocity and proper motion.  OCs may not be physically bound objects, but they form an easily-recognized ensemble.  Less well-defined are groups and associations, terms which are used interchangeably here.  The youngest stars we know of are found in star-forming regions, in close physical proximity to one another, but those groups are clearly too sparse to be bound.  At somewhat greater ages ($\sim10$ Myr), loose and very sparse groupings have recently been identified in the solar neighborhood \citep{Zuck04}.  Most of the bona fide OCs have ages from $\sim10$ to $\sim200$ Myr, older clusters being rare due to their tidal disruption in the Galaxy.  These sparse groupings are at least as much the subject of study for their astrophysical properties as they are benchmarks for age indicators.


\section{GENERAL COMMENTS}

\subsection{Attributes of a Useful Age Indicator}

We cannot measure stellar ages directly and so we use other tactics to address the problem.  
\citet{barnes07}
has discussed some of the attributes of a measurable quantity that would make it an ideal age indicator.  They include having a well-defined quantity that can be measured for single stars and that is sensitive to age but not other factors.  The quantity should be able to be calibrated and with known uncertainties.  The relation between the indicator and age should be invertible so that age can be calculated, and when applied to a coeval sample one should get consistent ages.  These criteria are all valid but address an ideal, and to that list should be added that the physics of the relation between the indicator and age is understood and that establishing the relation involves the fewest possible assumptions.  

There is no single ideal age indicator, and for real stars in the real Universe we have to settle for less, often much less.  Our errors may be dominated by systematic effects since we do not usually know what inherent physical relation to expect.  The sensitivity of the indicator may be useful for only a limited age range or mass range.  Inherent scatter in the quantity measured may mean that the age indicator is not very useful for single stars but can work if at least a few stars are being considered as an ensemble.  Clusters are ordinarily used to calibrate and to test how well a coeval population behaves, but open clusters more than $\sim1$ Gyr old are inherently rare and tend to be distant.

We may be willing to get by with much less.  In many cases it suffices to get the ordering of ages right, even if the age scale remains uncertain.  In other cases it may be enough to set limits to an age.  There are no perfect age indicators, and each decade of stellar age and each stellar type presents its own problems and has its own relevant indicators.  In order of usefulness, age indicators can be used, first to {\em classify} or {\em associate}, as is done, for instance, when a strong Li feature is used to confirm that a star is a member of the T Tauri class.  Second, an indicator can be used to {\em limit} an age, perhaps by imposing an upper bound.  Next, it can be helpful to {\em sort} ages in order to get the order right.  Finally, we can hope to {\em quantify} ages so as to specify a numeric age or age range.

\subsection{A Scheme for the Quality of Age Estimation Methods}

I break down the available age estimation methods into five levels of quality, based on how straightforward it is to go from the measurement to the age and how well we understand the physics at each step.  Each of the methods listed will be explicated further.  The table below summarizes my own view of how suitable the different methods are within different domains as a means of illustrating how the subjects will be presented.
\marginpar{\bf Table of methods: see separate PDF}

\subsubsection{FUNDAMENTAL AND SEMI-FUNDAMENTAL AGES}

An age is {\em fundamental} if the underlying physical processes are completely understood and all the needed observations are available.  The one and only fundamental age we have to work with is that of the Sun (see sidebar), and it is the result of measuring the decay products of long-lived isotopes in solar system material, something we can do for no other star.  

\marginpar{\bf Sidebar 3: The Sun as a Benchmark\\ (see p. \pageref{sidebar3})}

There are two stellar age estimation methods that can be considered {\em semi-fundamental} because they involve making only a few assumptions, those assumptions appear to be well-founded, and the assumptions do not influence the derived ages to a large degree.  The first, nucleocosmochronometry, applies to individual stars of the thick disk and halo and involves measurement of the decay of U or, especially, Th.  This is done for individual stars.  The method has been used only on metal-poor stars because it is only in those cases that the weak Th and U absorption lines can be measured in the presence of many blends.  The complex spectra of stars in and around the Th and U lines remains an important limitation on this method.

The second semi-fundamental technique can only be applied to a group of young stars and involves tracing the Galactic orbits of the stars back in time.  An assumption must be made about the Galactic potential through which the stars move, but that is not critical.  The age of the group is assumed to be when the stars were in closest physical proximity to one another.  This method of expansion ages works only for young groups because massive objects in the Galactic disk tidally disrupt groups and clusters on a time-scale of $\sim200$ Myr, leading to ``disk heating.''  If a group has an age less than $\sim50$ Myr we can be confident it has not yet been so disrupted.  Also, the uncertainties involved in calculating the stellar kinematics make it difficult to go back farther in time than $\sim30$ Myr.

\subsubsection{MODEL-DEPENDENT AGES}

The majority of the age estimation techniques used have built-in model dependencies.  The best-known method for clusters -- isochrone fitting -- starts with a series of calculated models and varies the properties of those models to achieve the best overall match to the observations.  In addition to the quality of the models themselves, there is also generally a calibration dependency in that \teff, for instance, must be related to an observed color or colors, and reddening taken into account since interstellar extinction is not a property of the models themselves.  Every step of the process adds a link to the chain.  Also, the process of isochrone fitting is itself a test of the models being used and our understanding of the underlying physics.  Examples of model-dependent methods include:

\begin{itemize}
\item Isochrone fitting, for clusters, or isochrone placement, for individual stars.  As we will describe below, for clusters isochrone fitting has the advantage of using a cluster's entire HRD, particularly key features that depend on specific physical processes.  This is what makes the HRD such a valuable tool for understanding stellar evolution, of course, and for testing models of stellar physics.  However, the quality of the isochrones depends on the completeness and accuracy of the models, and missing or inadequately understood processes can lead to systematic error.

\item Asteroseismology -- the detection of oscillation modes in stars -- is especially promising for single solar-type stars and for older stars because the low-order modes one can detect in an unresolved object pass through or near the star's center, making them an indicator of the star's central density, which is to say its age.  Interpreting asteroseismological observations requires detailed models, but the physics in those models is generally different than for MSTO stars in clusters and is well-understood, being for stars similar to the Sun.

\item The lithium depletion boundary (LDB) seen in very young open clusters is the point at the bottom of the main sequence where Li reappears.  Stars more massive than the LDB astrate all their Li in approaching the MS, while less-massive objects cannot reach the necessary internal temperature.  The observations needed to locate the LDB in a CMD are challenging because such objects are inherently very faint, but the underlying physics is fairly simple and well-understood.

\end{itemize}

\subsubsection{EMPIRICAL AGES}

Empirical ages are based on observed relations between a measured property and age.  The full physics of the relation is not understood and so there is no predictive power, but there are at least reasonable and plausible scenarios to explain what is seen.  Empirical ages must be empirically calibrated by measuring the relevant property in members of open clusters.  Because of the scarcity of old open clusters, combined with the difficulty of measuring the various properties in older stars, the Sun is often used to establish the relation for older stars and this adds an additional assumption.

The most direct of these empirical methods is what  \citet{barnes07} has called gyrochronology: using rotation periods to derive an age.  Less direct are methods using various forms of stellar activity (seen in the cores of the Ca {\sc ii} H and K  or the Mg {\sc ii} h and k lines, in H$\alpha$, and in x-rays, primarily).  Using activity is less direct because the activity appears to be dependent on the rotation.  In addition, activity diagnostics are inherently variable.  The last empirical method uses lithium abundances in F, G, and K stars.

\subsubsection{STATISTICAL AGES}

Two statistical correlations have been discussed extensively: the age-metallicity relation (AMR) and the net increase in Galactic space motions with time (``disk heating'').  It is not clear than an AMR relation actually exists in that the observed correlation may be due to an overall connection between Galactocentric radius and metallicity, with stars getting scattered into the solar neighborhood from the many places in our Galaxy where they formed.  At any rate, the correlation is at best a coarse one and is only suggestive for an individual star.  As a counterexample, there are recently-formed stars with metallicities about a factor of two below solar.

There is also a general trend of stellar space motions increasing with age, but note that both the Sun and the $\alpha$ Centauri system are $\sim4-5$ Gyr old yet have space motions near the Local Standard of Rest.  Again, a high space motion is suggestive of an old age, but it is also possible for individual younger stars to have been ejected from a cluster and for older stars to appear unaffected by these forces.  However, the net kinematic properties of, for instance, a volume-limited sample or another well-defined ensemble can be used to estimate a mean age.


\section{FUNDAMENTAL AND SEMI-FUNDAMENTAL AGES}

As noted, the one and only fundamental age in stellar astrophysics is that of the Sun, based on measuring abundances and daughter products of radioactive nuclides in meteoritic material.  Similar nuclide abundances can be measured in the spectra of stars, but without full knowledge of all the nuclides involved some critical assumptions must be made.  Those assumptions appear to be reasonably sound on both theoretical and practical grounds, leading to the first semi-fundamental technique, nucleocosmochronometry.

\subsection{Nucleocosmochronometry\label{nucleo}}

The ability to detect and measure Th and U in metal-poor stars has made nucleocosmochronometry particularly attractive for studying stars formed in the earliest epochs of the Galaxy and the processes that have enriched them.  Such stars are generally much too far away to have measured parallaxes that could lead to luminosity determinations and isochrone placement, but are compelling for understanding the time-scales of nucleosynthesis in the early Galaxy and for comparing what we believe to be the oldest stars to the ages of globular clusters \citep{vdb96}.

Nucleocosmochronometry derives stellar ages by measuring the decay of long-lived isotopes.  One seeks an isotope with a half-life comparable to the age of the object under study so that a measurable quantity remains; the isotope must offer detectable features; and the element's abundance should be dominated by that isotope.   The decay of these isotopes involves fully understood physical processes.  However, an initial abundance for the elements must be assumed because it cannot be measured directly.  This is ordinarily done by scaling from other r-process abundances and assuming production ratios of the elements being measured.  There remains significant uncertainty in these steps. Different calculations can lead to production ratios that differ by nearly a factor of two.

\citet{cowan1991} provided a thorough review of the history and early application of this method, and \citet{beers05} and \citet{frebel09} provide additional information.  \citet{ludwig09} present a summary of recent work on nucleocosmochronometry and the errors associated with it.  The two species used are $^{\rm 238}$U ($\tau_{\rm 1/2}=4.47$ Gyr) and $^{\rm 232}$Th ($\tau_{\rm 1/2} = 14.05$ Gyr).   Of the two, Th is more easily observed, with a Th {\sc ii} feature at 4019 \AA.  U is more difficult to detect except in some cases of highly-enhanced r-process abundances \citep{frebel07}.  Other nuclides with suitable half-lives have been sought that present good spectrum lines but none have been found.  

The U and Th isotope abundances are determined relative to other rare earths such as Eu, Os, Hf, or Ir.  Ideally the comparison element is one formed in just the same way as the Th, say, but this connection is an assumption that adds uncertainty.  Studies by \citet{cowan99} and others cited there tend to support the idea that r-process elements are formed in consistent proportions.  \citet{kratz07} argue that Hf should work best because it is so close in mass to Th, yet the analysis of \citet{ludwig09} does not bear that out, with Th/Hf ratios producing consistently unphysical results.  An illustrative case is provided by \citet{sneden2003} in their reanalysis of the very metal-poor star CS 22892-052.  They derived a net age of $12.8\pm3$ Gyr.  This is, of course, consistent with WMAP's age for the Universe of $13.7\pm0.2$ Gyr \citep{bennett03}, but the individual ages from various ratios in \citet{sneden2003} differed by nearly a factor of two (from 10.5 to 19.2 Gyr).  Another very metal-poor star with enhanced r-process elements, HE 1523--0901, presents a better case, having detectable U \citep{frebel07} and highly consistent ages from both U and Th, with an average of 13.2 Gyr.

The other examples of recent analyses cited in \citet{ludwig09} present much less consistency.  Often the calculated ages exceed 20 Gyr and only in a few cases are they below the WMAP age.  In a few cases the derived ages are $\sim1$ Gyr or less or even formally negative.  No one comparison element by itself provides better consistency.  The promise of nucleocosmochronometry to provide ages for, say, globular clusters that are independent of isochrone fitting will be fulfilled only with significant additional effort.  Systematic errors from different calculated production ratios should be able to be reduced, and the improving art of spectrum synthesis will lead to better results as well.  For comparison, about a half century ago \citet{fowler60} determined an age for the Galaxy of $15\pm3$ Gyr from nuclear analysis of solar system material.

In principle nucleocosmochronometry would be ideal for old main sequence stars, where few other age estimation techniques are useful, but in practice this method can generally be applied only to metal-poor stars because of the difficulty of measuring the U or Th feature in solar-metallicity stars in the presence of the many blending features.  In addition, the r-process elements in a star like the Sun will have come from many separate events, while in the oldest stars formed early in the Galaxy's history fewer nucleosynthesis events will have taken place before the star formed.  Also, in some cases the comparison element will have a contribution from the s-process as well as the r-process.  Nevertheless, the Sun's age has been calculated as a test of the procedures.  \citet{ludwig09} use the measurements of \citet{caffau08} to derive solar ages from 1.7 to 22.3 Gyr.  The Th/Eu ratio appears to  yield the most reasonable results, but with poor precision.

Th/Eu ratios have also been used by \citet{delP05b}, \citet{delP05a}, and \citet{delP05c} to determine an age for the Galactic disk of $9\pm2$ Gyr.  In this case the age is not from individual stars but from matching the observations to models of Galactic chemical evolution that take several effects into account that enrich and destroy Th and Eu with time.

\subsubsection{SUMMARY OF NUCLEOCOSMOCHRONOMETRY}

\begin{description}
\item[$+$] The method is applicable to the Galaxy's oldest stars, independent of distance.
\item[$+$] The method is applicable to individual stars.
\item[$+$] The physics of the process is well-understood.
\item[$-$] Spectra of high resolution and high signal-to-noise are required to detect the weak features, limiting the method to brighter stars.  Generally only a single feature of an element is detected.
\item[$-$] The method works only for low-metallicity stars because of line blending.  Very few stars have been found with detectable Th or U features.
\item[$-$] The uncertainties are at least 20\%.  \citet{ludwig09} estimate that with even favorable errors that the uncertainty in age for an individual star is about 2.5 Gyr.
\item[$-$] Significant systematic errors remain, particularly in the expected production ratios of rare earth elements.
\end{description}

\subsection{Kinematic or Expansion Ages}

Kinematic ages are determined by tracing the motions of a group of stars into the past to determine when they were in closest physical proximity, which is assumed to be the time of formation.  Alternatively, one can plot proper motion in declination, say, against declination to get a linear expansion rate, which then implies an age.  The concept goes back to work on the expansion of OB associations by Ambartsumian and Blaauw around 1950 (see, e.g., \citet{blaauw64}), although the early results were inconsistent and sometimes in conflict with evolutionary ages for massive stars.  A related age-dating method connects runaway stars back to their point of formation and assumes linear motion.  That topic will not be discussed here as runaway stars are important in their own right as tracers of dynamical processes.

The attraction of the kinematic method, which obviously can only be applied to a group of stars, is that it is independent of any stellar physics since it relies predominantly on astrometry (plus radial velocities).  However, there are two major limitations, one practical and one fundamental.  The practical problem is that applying the method requires high-quality kinematic data in all three dimensions (i.e., including parallax and radial velocity).  Attempts have been made to use only proper motions because they can be very precise, but the simulations of \citet{brown97} show that doing so leads to consistently underestimated ages and overestimates of the initial sizes of the star-forming regions.  Also, the resultant kinematic ages are nearly always significantly less than evolutionary ages \citep{brown97}, for reasons that are not well understood.

Even with full 3-D kinematic data, good accuracy is required to get useful results and such data are still not available for many groups of interest because they are too distant.  However, there are a number of groups of very young stars near to the Sun (see \citet{Zuck04}) that have members with {\it Hipparcos} astrometry.  Kinematic ages for these groups are summarized in \citet{makarov07} and \citet{Fern08}.  In most cases the kinematic ages are greater than ages from pre-main sequence isochrones, but, as discussed below (Sec. \ref{PMS}), the isochrones for PMS stars have significant uncertainties.  The errors in the measured input quantities are such that going back further than $\sim20-30$ Myr is problematic with the kinematic technique.  Another limitation is that some young groups appear to have little net expansion (\citet{mamajek05}; \citet{makarov07}), so that a time when the members were closest together is ill-defined. 

The fundamental limitation on kinematic ages is that stars and clusters encounter massive objects as they orbit in our Galaxy.  These objects exert forces that disrupt clusters and stellar orbits, creating field stars and leading to ``disk heating,'' which is the observed increase of vertical scale height with age \citep{wielen77}.  The time-scale for these encounters is $\sim200$ Myr \citep{janes88, janes94}, roughly the Galactic rotation period.  Thus one can be reasonably confident that the Galactic motions of young stars are undisturbed if they are $<10^8$ Myr old, but for older stars Galactic orbits cannot be reliably traced back in time.

\subsubsection{SUMMARY OF KINEMATIC AGES}

\begin{description}
\item[$+$] Kinematic age estimation uses few assumptions and involves no stellar modeling.  The only requirements are good three-dimensional space motions, meaning high-quality astrometry and radial velocities.
\item[$+$] The method is applicable to very young ensembles of stars for which few or no other age estimators work well.
\item[$+$] The kinematic method can be applied to stars in any mass range.
\item[$-$]Kinematic traceback can only be done on an ensemble of objects, not individual stars.
\item[$-$]The method works only for young groups, up to $\tau\sim20$ Myr.
\item[$-$] Accurate parallaxes in particular are required, which are often unavailable for faint, low-mass stars of interest.
\end{description}

\subsection{The Lithium Depletion Boundary}

LDB ages have been determined for only a very small number of nearby clusters that have been age-dated in other ways.  The significance of LDB ages is for the overall OC age scale and so a discussion of the lithium depletion boundary is given in Section \ref{LDB} in the presentation on OC ages.


\section{MODEL-DEPENDENT AGES FOR INDIVIDUAL STARS}

For the most part, the stellar models used to estimate the ages of individual stars are the same ones that go into determining OC ages, but there are different ways those models are applied because less information is available for the single star.  OC ages and how they are used to calibrate star ages are discussed in Section \ref{OC-age}.

\subsection{Isochrone Placement for Main Sequence Stars}\label{isochrone-MS}

The principle of deriving an age from placing a star on model isochrones in the HRD is straightforward, but in practice it is difficult and problematic.  One starts with the same isochrones used to fit to cluster CMDs, based on well-tested stellar models.  However, in fitting a cluster one has not only more stars to improve a fit statistically, those stars are also distributed in mass, and it is the full behavior of the models over that mass range that is being fit, leading to a generally well-constrained situation, depending, of course, on the accuracy of the input quantities and the completeness of the physical processes that are included.

For single stars, that distributed fit is not possible.  Moreover, isochrone shapes are complex, particularly near the MSTO, and multiple isochrones can pass through a given point.  Some of that degeneracy is removed if the abundances in the star are known, and additional constraints exist if the star is in a binary system.  The contours and degeneracy of isochrones means that the location of a star in a CMD cannot be simply inverted to yield an age.  They also mean that extracting a most-probable age and realistic uncertainties require special analysis.  This can possibly result in a well-defined age, but sometimes only a limit can be established or there is little constraint at all.

The method starts with best estimates of temperature, metallicity, and luminosity, together with uncertainties.    \citet{edvardsson93}, \citet{ngbertelli98}, \citet{nordstrom04}, and \citet{holmberg09} have applied this method to the Geneva-Copenhagen survey of $\sim14,000$ nearby FGK stars.  One can interpolate among the model isochrones to the location of the star, but this introduces biases both because of the complex shapes of isochrones and their non-uniform spacing. In particular, input uncertainties in \teff\ and \Lbol\ may be gaussian, but the uneven spacing of isochrones near the ZAMS (with an accelerating increase in \Lbol\ as a star ages) leads to equally probably young- and old ages from naive placement of the star, when in fact most stars should fall in the dense ZAMS region; this effect biases results toward greater ages and is essentially the same as the Malmquist or Eddington bias seen in other areas of astronomy.  This bias in resultant ages is exacerbated by the fact that stars on and near the ZAMS often have indeterminate ages and the more evolved stars have well-defined ages (see below).  This can also bias apparent ages to higher values.   This bias can be seen in one of the earliest efforts at precision isochrone placement \citep{perrin77}, where most derived ages of disk stars exceeded 10 Gyr and ranged up to $\sim25$ Gyr.

Some recent studies have continued to estimate ages from direct isochrone interpolation (e.g., \citet{karatas05}), but more sophisticated techniques have been applied to attempt to overcome these difficulties.  \citet{ponteyer04}, for example, apply bayesian methods to the F and early-G dwarfs in the Geneva-Copenhagen sample, in particular stars with $\tau \approx$3-15 Gyr, which is to say stars that have evolved for at least $\sim1/3$ of their MS lifetime so that they have moved away from the ZAMS.  \citet{ponteyer04} were able to lessen biases in the ages, although individual ages had large uncertainties.  \citet{jorgensen05} expanded on \citet{ponteyer04} by starting with a large sample of synthetic stars with assigned masses and ages.  After adding errors they tested their ability to recover the starting ages.  They tried several techniques and found bayesian methods gave the most accurate net answer with the least error.  That uncertainty was about 0.20 dex, or 50\%, on average.  Relatively massive and evolved stars could be dated to within about 20\%.  
Note that their sample did not go as low in mass as 1 \MS, nor did they treat ZAMS stars.  Their interest was in improving fits to the MSTO in sparse older clusters and they were able to do so for with remarkably small fitting errors.

\begin{figure}[t]
\includegraphics[scale=0.50]{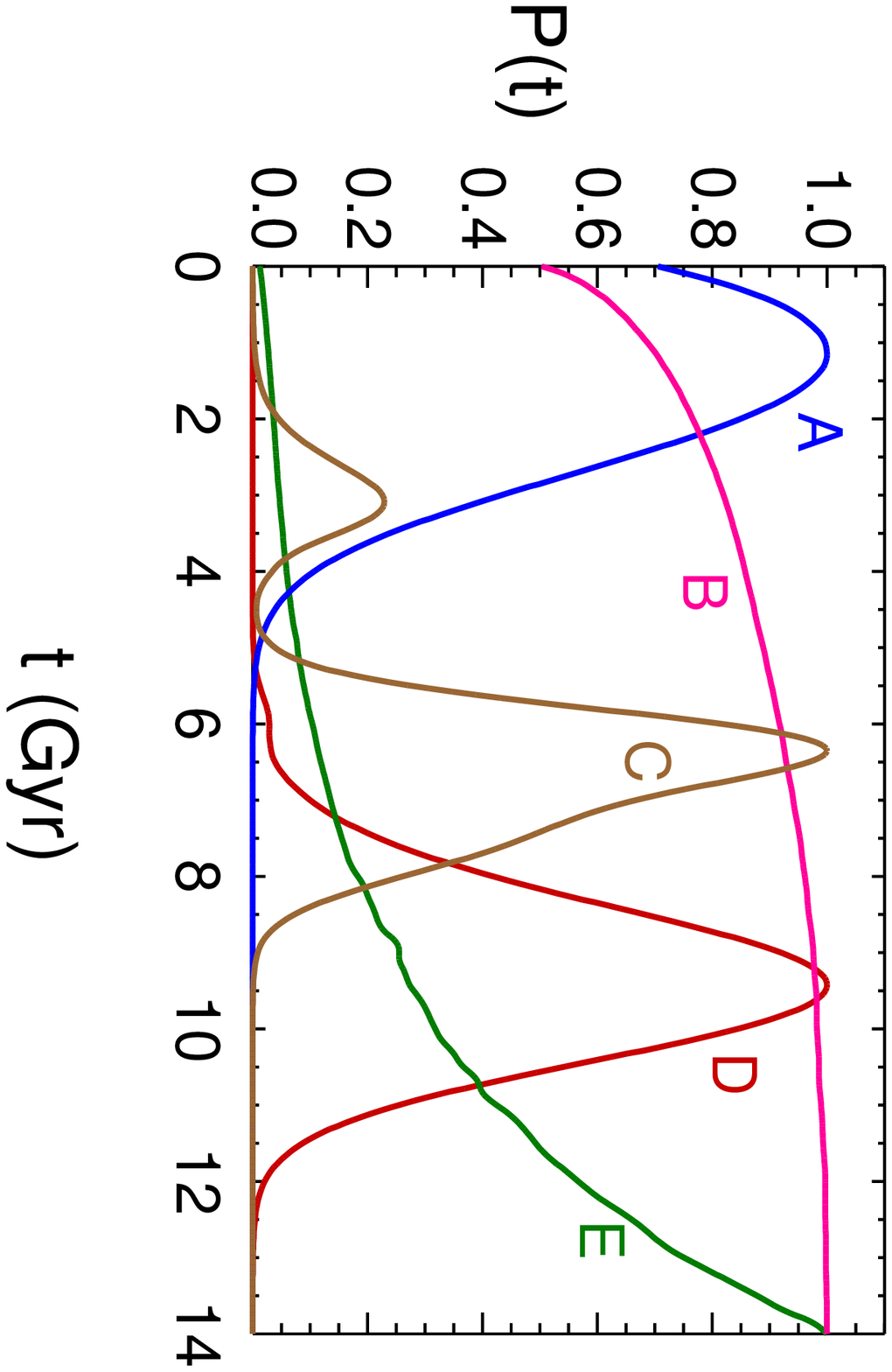}
\caption{\label{Takeda-PDF}
Normalized age probability distribution functions (PDFs) for isochrone placement of individual main sequence solar-type stars, chosen to show representative cases (adapted from \citet{takeda07}).  Stars C (brown) and D (red) show ``well-defined'' ages, having a single or at least a dominant peak.  For star A (blue), an upper bound to age (2.9 Gyr) results plus a best-estimate age (1.2 Gyr).  Only the lower-bound for star E (green) can be specified (12.2 Gyr).  No meaningful age can be derived for star B (pink).
Of the stars studied by \citet{takeda07}, 2/3 had well-defined ages (in the sense they describe, with a well-delineated peak in the PDF) and of those only 25\% had errors in age of 1 Gyr or less.
}
\end{figure}

\citet{takeda07} also developed a bayesian method that they applied to a large sample of G dwarfs with masses from about 0.8 to 1.3 \MS, significantly lower masses than what either \citet{ponteyer04} or \citet{jorgensen05} considered.  Figure \ref{Takeda-PDF}, 
\marginpar{\bf Fig. \ref{Takeda-PDF}: Probability distribution functions of age from Takeda et al.}
taken from \citet{takeda07}, illustrates important aspects of the technique in that well-defined ages result in only some circumstances, while in other cases only limits to age can be determined.  The net effect of this unavoidable effect, as noted, is to add a bias to the resulting age distributions, favoring older ages over younger, and more massive stars over the less massive. 


Derived uncertainties in ages from isochrone placement are significant.  In their analysis of their synthetic dataset, \citet{jorgensen05} were able to recover ages to better than 20\% only for relatively massive and well-evolved stars.  Less-evolved stars had ages good to about 50\%, and stars near the ZAMS were indeterminate.  

\citet{takeda07} got similar uncertainties, with more massive stars ($\sim1.4$\MS) having ages good to $\sim10$\%.   In the \citet{takeda07} study, which was able to use some of the highest quality data available for FGK stars, only 2/3 of the stars ended up with a well-defined age in the sense that a clear peak was present above the 60\% probability level.  Of those stars with well-defined ages, only 25\% had formal errors of less than 1 Gyr, and those are fitting errors that do not take into account systematic effects in the models.

The age errors are dominated by uncertainty in \teff, usually taken to be 80-100 K.  Given a good parallax, the only factor adding to uncertainty in \Lbol\ is the bolometric correction, and BCs are well-established, at least for stars similar to the Sun.  But the stellar temperature scale remains fundamentally uncertain, in part because we try to characterize what are manifestly non-uniform stellar atmospheres with a single parameter that describes the total energy output (\teff) but which does not necessarily describe the temperature structure of the star's atmosphere and the conditions under which spectrum lines are formed.  Better ways of describing the atmospheres of late-type stars are needed in order to better relate observations to models.

\subsubsection{SUMMARY OF ISOCHRONE PLACEMENT FOR MS STARS}

\begin{description}
\item[$+$] Isochrone placement uses models that are based on well-understood stellar physics, especially for stars similar to the Sun.
\item[$+$] The method is well-suited to older or more massive stars where other techniques (such as the empirical methods) are of little use.
\item[$-$] Not all stars end up with an isochrone placement age even with good quality observational data.  This effect can bias interpretations of the ages of a group of stars.
\item[$-$] Unresolved binaries will appear to be more luminous (i.e., older) than they really are.
\item[$-$] Error analysis is complex and non-gaussian.
\item[$-$] Uncertainties are significant even when a well-defined age results, often being 20--50\% before systematic effects are taken into account.
\item[$-$] Isochrone placement is limited to stars at least as old as $\sim1/3$ of their overall MS lifetime so that they have moved away from the ZAMS in the CMD.  This favors more massive stars.
\item[$-$] Ages and uncertainties are limited significantly by errors in \teff.  Incomplete knowledge of composition is also a limitation but generally a lesser one if a spectroscopic metallicity is available.  The star's helium abundance is crucial, of course, but with little information to work with He is assumed to be solar.
\item[$-$] Many different isochrones can pass through a given location of the HRD.  Resolving this degeneracy requires at least good knowledge of the star's abundances.
\end{description}

\subsection{Isochrone Placement for Pre-Main Sequence Stars}\label{PMS}

At the most basic level, for PMS stars it is very difficult to be sure one has determined an accurate \Lbol\ and \teff, both of which are essential for placing a star in an HRD.  Star-forming regions remain rich in the residual material from the formation process, leaving with it ambiguity about what reddening law to apply, how much overall extinction may exist, and how much local extinction may play a role, especially since the material closest to stars is clearly not  isotropically distributed.  In some cases it can be difficult to isolate the light of a given star from surrounding material or other objects in close proximity.

Placing PMS stars on isochrones is generally done for groups of stars that are found in a given star-forming region or in an ensemble.  Nevertheless, each star must be treated individually because of the many factors that affect its location in the HRD.  Reddening and extinction affect both colors and magnitudes, and both can vary significantly in a star-forming region because of the residual material.  PMS stars are inherently variable too.  Even after correcting for reddening, the colors of PMS stars are not reliable temperature indicators.  Some PMS stars exhibit infrared excesses, indicative of circumstellar material.  Others show blue excesses and/or excess continuum emission (``veiling'') that is attributed to accretion.  \teff\ values for PMS stars are best assigned from spectral types and that requires additional observations, and the result are imprecise.  The relation between spectral type and \teff\ is likely different for PMS stars than for dwarfs, in part due to gravity effects \citep{luhman03} and also because PMS stars appear to have highly inhomogeneous atmospheres not well characterized by a single temperature.

\begin{figure}[t]
\includegraphics[scale=0.50]{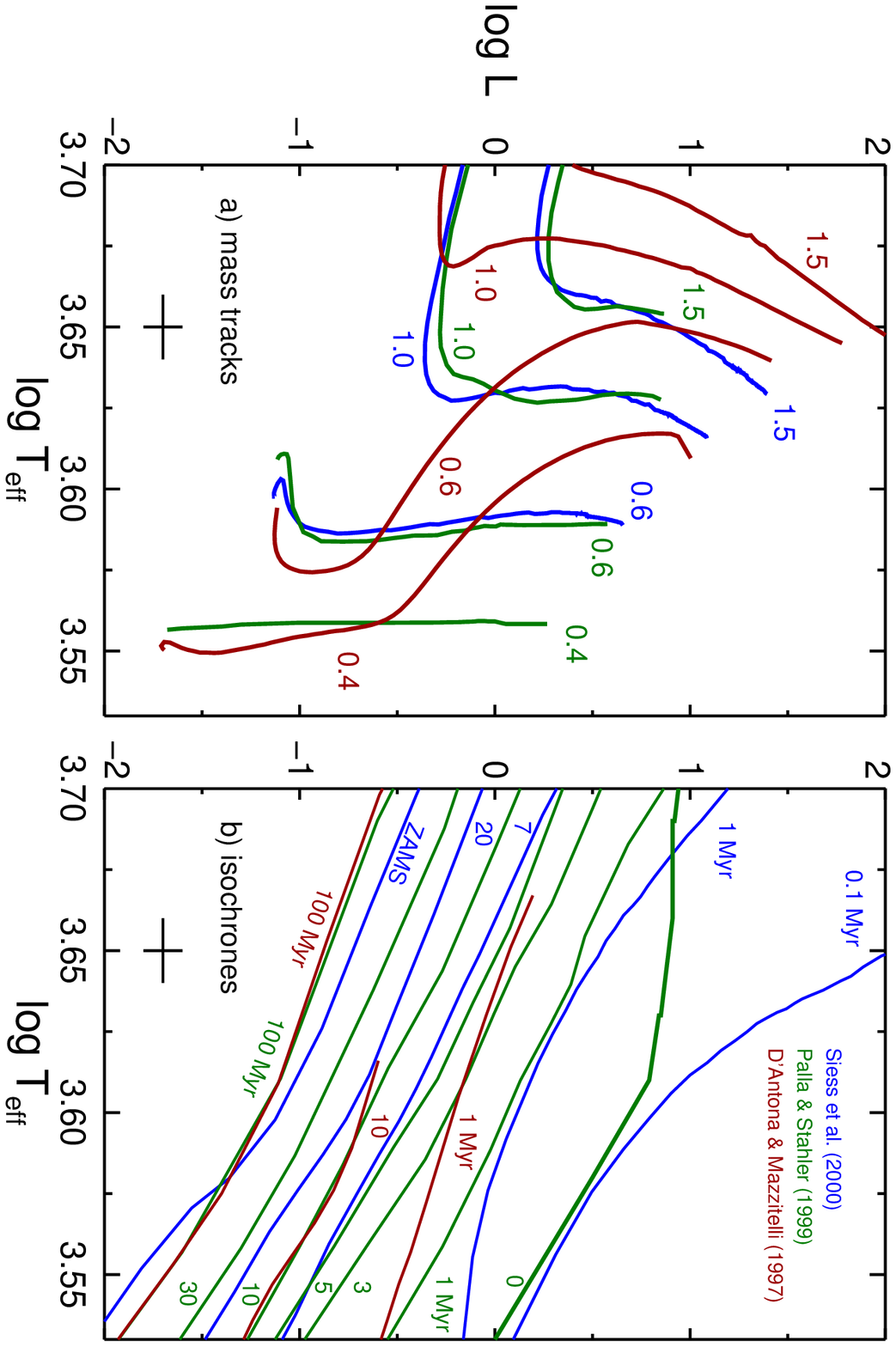}
\caption{\label{PMS-isochrones}
A comparison of several sets of PMS models.  For both panels the blue lines show the same mass tracks and isochrones from \citet{siess00} that are shown in Figure \ref{ONC-HRD}, except for a more limited range of mass.  Three sets of models are shown: \citet{siess00}, blue; \citet{palla99}, green; and \citet{DM97}, red.  The three sets do not always cover the same masses or ages, and so a representative subset is shown.  Mass tracks are shown on the left for four masses: 0.4, 0.6, 1.0, and 1.5 \MS.  The cross near the bottom shows the effect of an uncertainty of $\pm100$ K in \teff\ and $\pm0.1$ dex in \loglbol.  Actual uncertainties in \teff\ for PMS stars are larger (200-300 K) and also larger in \Lbol\ (see Fig. \ref{ONC-HRD}).  The right panel shows isochrones, marked with ages in Myr.  The ``0 Myr'' isochrone of \citet{palla99} is their ``birthline'' where stars first become visible.  Both sets of mass tracks and isochrones are broadly similar in many cases but deviate significantly at the youngest and oldest ages shown.  The different nature of the mass tracks of \citet{DM97} is especially striking.  Note the considerable uncertainty in age or mass that results from the models, independent of observational uncertainty.
}
\end{figure}

Even if one is able to work through these problems to some satisfactory level, PMS ages are determined from comparing the locations of stars in an HRD to calculated isochrones and mass tracks.  These models differ substantially, mainly because of differing treatments of convection and also the equations of state and the radiative opacities that are used.  Uncertainties in the ages of factors of 2-3 are likely.  Figure \ref{PMS-isochrones} 
\marginpar{\bf Fig. \ref{PMS-isochrones}: Examples of PMS isochrones.}
shows examples of calculated PMS isochrones for low-mass stars, with the differences growing in going to the cooler K and M stars which form the dominant portion of the T Tauri population.

\begin{figure}[t]
\includegraphics[scale=0.50]{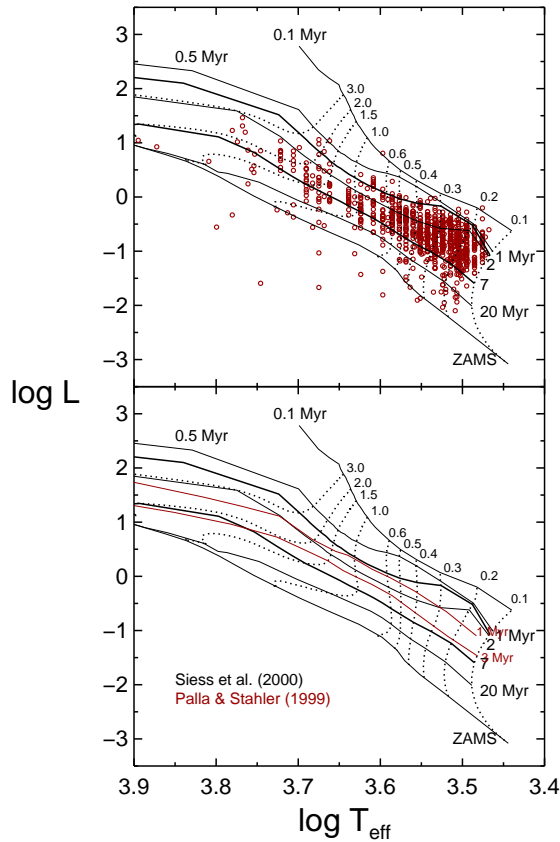}
\caption{\label{ONC-HRD}
An HRD for the Orion Nebula Cluster (ONC), adapted from \citet{dario09}.  The ONC is one of the most-observed clusters of PMS stars and so illustrates the difficulties involved in studying the ages of very young low-mass stars (the highest-mass stars in the ONC are not shown).  Temperatures were determined from spectral types (mostly from \citet{hillenbrand97}) to avoid the effects on the photometry of reddening, veiling, accretion, and so on; this accounts for the quantization of most of the \teff\ values.  Luminosities were determined from optical $UBVRI$ photometry after correcting for reddening and veiling.  The $R_V$ value assumed for the reddening is 3.1, and the assumed distance to the ONC is 414 pc, as measured by radio interferometry.  Individual values of reddening for each star were calculated using the multi-color photometry.
The evolutionary tracks and isochrones shown are from \citet{siess00}.  The evolutionary tracks (dotted lines) have their masses shown along the 0.1 Myr isochrones, from 0.1 to 3.0 \MS.  Several representative isochrones (solid lines) are plotted for 0.1, 0.5, 1, 2, 7, and 20 Myr, plus a ZAMS.  The 1 and 7 Myr isochrones are bolder.
The ages of these ONC stars scatter in \Lbol\ enough to appear to range from $\sim1-10$ Myr in age, peaking at $\sim3$ Myr.  Despite this considerable spread in luminosity, \citet{dario09} did not conclude that a real age spread was clearly present because of the uncertainty added by stellar variability and scattered light from circumstellar material.
Different sets of isochrones yield different results, both in absolute age and in the spread.  For example, the evolutionary models of \citet{palla99} lead to a mean age of about 2 Myr but with the age depending on mass (see \citet{dario09}).
}
\end{figure}

The classic case study of a star-forming region is that of the Orion Nebula Cluster (ONC).  \citet{hillenbrand97} presented a thorough analysis of the available photometry and spectroscopy that has recently been updated with {\it HST} photometry for many more stars by \citet{dario09}.  \citet{dario09} present an analysis that statistically takes into account reddening, accretion effects, and so on, plus average luminosities for well-observed stars, yet there remains a significant spread in \Lbol\ at a given \teff.  These spreads are typically seen in all star-forming regions and are not fully understood.
\marginpar{\bf Fig. \ref{ONC-HRD}: HRD for the Orion Nebula Cluster.}

A good test of isochrone placement is offered by binary systems when consistent data of good quality are available for both components.   \citet{kraus09} found that binaries in Taurus-Auriga, when placed on PMS isochrones, differed significantly less in apparent age than did randomly-selected pairs of stars from that star-forming region.  This was interpreted to mean that stars within binaries form with less delay relative to one another compared to the overall time-scale for formation within Tau-Aur.  Also, the binary components had relative \teff\ and \Lbol\ values consistent with the isochrones used, a blend of models from \citet{baraffe98, chabrier00} and \citet{DM97}.

However, one can still hope to determine if significant age {\em spreads} exist within star-forming regions; see Sec. \ref{age-spread}.  

Measured masses from PMS binary star orbits are rare and imprecise, yet are essential for calibrating the models.  Even the nearest star-forming regions are too far away to have trigonometrically-determined distances, although in a few cases (Orion and T Tauri itself, notably), radio interferometry has been able to greatly improve on that vital quantity.

As stars approach the ZAMS, their evolution slows and the isochrones crowd close together, too close relative to error bars we now have on luminosities, making age estimation difficult even if the models are accepted at face value.  A few ensembles of co-moving young stars have been identified \citep{Zuck04} and have members with trigonometric parallaxes from {\it Hipparcos}.  The ages of these ensembles have been estimated from isochrone fitting which works, as for star-forming regions, because a number of stars are present.  In a very few cases (e.g., \citet{boden05}) it is possible to estimate the age of a single object, in part because the ``object'' is in fact a multiple system, but even then the age is imprecise.

Systematic differences between sets of model isochrones for PMS stars range from 20\% for the hotter stars (F and G stars, with better-understood physics) to as much as factors of 4-5 for late-K and M stars.  Aside from that, observational uncertainty dominates in comparing stars within star-forming regions (to determine age spreads) or between different regions.

\subsubsection{SUMMARY OF ISOCHRONE PLACEMENT FOR PMS STARS}

\begin{description}
\item[$+$] For PMS stars, isochrone placement uses a property of the stars that is changing rapidly and systematically, the luminosity.  However, the luminosity may not be changing monotonically, given the many factors that influence it.
\item[$+$] The method is potentially well-suited to more massive stars because differences in the models are less.
\item[$-$] Differences in the available models can be very large, leading to significant systematic uncertainty.
\item[$-$] Placing individual stars in the HRD requires estimates of \teff\ and \Lbol, both of which are affected by non-uniform reddening and extinction, and by circumstellar material or accretion.
\item[$-$] Unresolved binaries will appear to be more luminous (i.e., younger) than they really are and will add to an apparent age spread.
\item[$-$] PMS models depend on assumptions of the accretion histories of objects that are difficult to verify.
\item[$-$] Error analysis is difficult because of many poorly-understood factors that lead from the observations to the modeled quantities.
\item[$-$] Isochrone placement for PMS stars is less certain as stars approach the ZAMS because isochrones get closer together.
\item[$-$] Ages and uncertainties are limited significantly by errors in \teff.
\end{description}

\subsubsection{AGE SPREADS IN STAR-FORMING REGIONS}\label{age-spread}

Even if the absolute age of a star-forming region such as the Orion Nebular Cluster may be difficult to establish, we would like to be able to determine the extent, if any, of age spreads within such groups.  This amounts to having a reliable age ordering scheme.  The usual method is to examine the spread in \Lbol\ and to compare that to expectations from known error sources to see if a residual spread remains that is attributable to an age spread.  The Orion Nebula Cluster (ONC) and the region around it presents a favorable case observationally in that the ONC is among the youngest of nearby star-forming regions, helping to make the effects age differences relatively larger.  The most recent analysis \citep{dario09}, like those before it, finds no convincing evidence for a spread in any one region (see Fig. \ref{ONC-HRD}).  There is, however, a progression in age spatially such that the outer regions of the ONC are older than the central regions \citep{dario09}.  \citet{jeffries07} studied NGC 2169, a 9 Myr old cluster, and found $\Delta \tau < 2.5$ Myr.

There must be a finite duration over which star formation occurs in a given region, the question is measuring it in the presence of many other effects.  One imagines, though, that binary systems would have very little lag in the formation of the two stars and so that binaries in star-forming regions might offer a means of assessing the apparent spread within a population with $\Delta \tau \approx 0$.  That expectation is borne out for the most part \citep{kraus09}, although inevitably some binaries are conspicuous exceptions for reasons that should help us understand the HRDs of very young groups of stars.

\citet{palla05} have applied another means of estimating the age spread within the ONC: differences in lithium abundances.  Li in PMS stars presents its own problems (see Sec. \ref{Li-depletion}), but again differences should be able to be established more reliably.  \citep{palla05} find several low-mass ($\sim0.4$ \MS) stars that appear to be genuine ONC members yet have depleted enough Li to be about 10 Myr old, an age much greater than the rest of the cluster ($\sim1-3$ Myr).

Binary systems also can provide consistency checks on methods of estimating ages and age spreads.  \citet{white05} show that a probable member of the Taurus-Auriga association lacks Li in both components, suggesting an age inconsistent with isochrones.  \citet{hartigan03} also looked at binaries in Taurus-Auriga and found that primaries appeared to be systematically older than their secondaries for three different sets of PMS evolutionary tracks and that some pairs were unusually discrepant, perhaps because of underestimated veiling.

The accretion histories of stars will influence their apparent ages and apparent age differences as well, and this has been addressed by \citet{froebrich06} and \citet{baraffe09}.  In particular, \citet{baraffe09} find that allowing episodic accretion can produce enough spread in \Lbol\ at the age of the ONC ($\sim1$ Myr old) to show an apparent spread of $\sim10$ Myr.  This episodic accretion also affects stellar radii and, possibly, the existence of the stellar birthline of \citet{stahler88}.

Testing for age spreads within star-forming regions is an instance of determining relative ages.  \citet{hillwhite08} present a review of this subject.

\subsection{Asteroseismology}

Stellar pulsations have long been a means of testing models of stellar physics because pulsations probe a star's interior.  Isochrone fitting works with the same models over the full extent of the HRD but has only the simple surface descriptors -- \Lbol\ and \teff\ -- to work with.  By contrast, helioseismology has provided such a wealth of detailed information and precise physical constraints on the solar interior that it has led to fundamental ramifications for particle physics.  

Nearly all types of stars pulsate to some extent; here I concentrate on solar-type stars because the oscillations provide a way of constraining a star's age, and I focus on just the parts relevant to ages since asteroseismology is itself a deep subject worthy of a penetrating examination.  See \citet{cunha07} for a recent review of asteroseismology and the physics underlying it.

For the Sun, $\sim10^5$ modes have been detected, but we cannot resolve the surfaces of stars and so are limited to modes of very low degree.  However, those low-order modes probe the deep interior of the star and so set valuable constraints on important properties, including, indirectly, the age.  The observations needed to detect oscillations in stars like the Sun are extraordinarily demanding.  The Sun's amplitude is $\sim1$ m s$^{-1}$ in velocity \citep{kjeldsen08a} and $4\times 10^{-6}$ in intensity \citep{lebreton09}, and these scale roughly in proportion to \Lbol\ and inversely with mass.   This degree of photometric precision is only achieved from space, but ground-based radial velocity observations have yielded results for a small number of the brightest FGK dwarfs.

For the Sun, the primary period seen is $\sim5$ min, but the power spectrum shows a pattern, with a large frequency spacing $\Delta \nu \approx 100 \mu$Hz and a small frequency spacing $\delta \nu \approx 10 \mu$Hz.   $\Delta \nu$ scales as the mean density of the star and is mainly sensitive to stellar mass.  $\delta \nu$ is sensitive to the gradient of the sound speed which reflects the star's evolution.  A plot of $\delta \nu / \Delta \nu$ versus $\Delta \nu$ has been shown to be a good diagnostic of mass and age for solar-type stars that should lead to age accuracies of $\sim10$\% (\citet{oti05}; \citet{mazumdar05}; \citet{kjeldsen08b}).  Detecting these frequency spacings requires a near-continuous data stream for about a week, meaning, when done from the ground, a multi-site campaign to eliminate diurnal aliases in the power spectrum.  At the same time, the signal sought is by its nature periodic and the mode lifetimes are long so that summing observations can yield the needed signal-to-noise.

Ages from asteroseismology use the same models applied to isochrone placement and so have the same model vulnerabilities.  As for isochrone placement, it is important to constrain basic stellar parameters as well as possible, particularly the composition.  In some cases it may be possible to independently measure a star's radius (via interferometry), mass (from a binary orbit), or mean density (from a planetary transit).  The great value of asteroseismology is that it yields more significant constraints on the models, and can include the He abundance and the size of the convective envelope or convective core \citep{lebreton09}.

Because the observations are so resource-intensive only a small number of stars have been observed (I am aware of five with asteroseismological ages; see, e.g., \citet{vauclair09}).  In addition, a full fit to the observations for any one star requires a custom-built stellar model tuned to each particular case.  As discussed below (Sec. \ref{future}), this is about to change significantly because of the {\it CoRoT} and, especially, {\it Kepler} missions, which will produce high-quality asteroseismological signatures for thousands of stars.  What has been a boutique business is about to go past retail to wholesale.  Among the few stars with detected seismic frequencies, even fewer have had ages determined.  However, the ages can be very precise, and, since they use information from much of the star, are likely to be more accurate as well.

With a sufficiently long time-span (months to a year), asteroseismology is capable of detecting aspects of the star's interior that arise from convection, which is to say the depth of the convective envelope.  Only {\it Kepler} is likely to yield such data, but it can provide for the first time critical parameters of interior structure for convective stars other than the Sun.

\subsubsection{SUMMARY OF AGES FROM ASTEROSEISMOLOGY}

\begin{description}
\item[$+$] Asteroseismology provides physical constraints for the star observed that are available in no other way.
\item[$+$] Asteroseismology is especially effective for older MS stars because the increase of He in the core increases the density, leading to an increase in the sound speed, which in reflected in the oscillation frequencies.  It is these stars for which the ages can be so intractable otherwise.
\item[$+$] The ages that result from a full modeling effort are probably accurate to $\sim10$\%, significantly better than any other age-dating method.
\item[$+$] Asteroseismology can be applied to individual stars at any evolutionary state.
\item[$-$] Asteroseismology uses the same stellar models used for isochrone placement (for individual stars) and isochrone fitting (for clusters) and has the same limitations.  At the same time, the stars observed are very similar to the Sun and so the models and the underlying physics should be sound.  Efforts are underway to better understand the limitations and problems in the models \citep{lebreton08}.
\item[$-$] Obtaining the needed observations for a particular field star is extremely resource intensive, requiring the largest telescopes and best spectrographs in a coordinated campaign, followed by modeling that is tailored to the source.
\item[$-$] Our understanding of the physics of PMS and low-mass stars is not yet sufficient to predict or analyze their oscillation properties.
\end{description}


\section{EMPIRICAL AGE RELATIONS}

Empirical age relations provide an observed change of a quantity with age that can be calibrated with clusters or stars that have ages determined in other ways.  Also, a plausible physical scenario exists to provide a framework for the observations, even if we do not fully understand the physical processes and cannot predict the behavior of the property.  Because empirical indicators are calibrated against objects with model-dependent ages (clusters mostly), they are secondary to the model-dependent methods.

The empirical relations discussed here apply to low-mass stars (main sequence and PMS) and this is because they all have their origin in the presence of a convective zone (CZ) in such stars.  In such stars, convection and rotation (especially latitudinal differential rotation) interact to create complex motions, and the material in the CZ is ionized and therefore conducting.  These motions can then regenerate a seed magnetic field to reinforce and amplify it: this is the dynamo mechanism.  Going further, we observe the Sun to have an ionized wind, and that enables the magnetic field to force corotation of the wind to well beyond the Sun's surface, leading to angular momentum (AM) loss.  Thus in this picture the gradual spindown of a star like the Sun is inevitable.  Also, stars that start with higher-than-average rotation rates produce stronger magnetic fields and so lose AM faster.  In other words, there is a feedback mechanism that leads to an initial spread in rotation rates among a coeval population converging over time.  

The magnetic fields that are generated through the dynamo mechanism manifest themselves as activity in a variety of forms.  Non-thermal energy is deposited into the chromosphere and corona.  Properties of a star's chromosphere can be observed by obtaining spectra in the cores of very strong absorption features such as H$\alpha$, the Mg {\sc ii} h and k lines, the Ca {\sc ii} H and K lines, or the Ca {\sc ii} infrared triplet.  At those wavelengths the stellar photosphere is suppressed by the absorption and the weak chromosphere can be seen.  Stellar coronae are seen as soft x-ray emission.

This scenario forms the paradigm that guides the study of spindown, the decline of activity, and the relation between the two in late-type stars.  It is important to emphasize that this picture is an hypothesis that has been partially tested but not completely.  The processes that feed energy into the chromosphere and corona are not understood even for the Sun.  However, this scenario is consistent with the observations available so far (but see a comment below).  Another consequence of the CZ is, we believe, the depletion of a star's surface Li abundance over time and that provides an additional empirical age indicator.  The spots and magnetic activity that are associated with these processes lead to photometric variability as well, and that is briefly described.

This subject of empirical age indicators has its own history, but the papers by \citet{kraft67} and \citet{skumanich72} have inspired much that followed.  \citet{skumanich72} used rotation, activity, and Li data then available for the few nearest OCs and for the Sun and noted that rotation and activity both appeared to decline in proportion to $\tau^{-1/2}$, while Li depletion followed an exponential.  The rotation data available at the time was from \vsini\ measurements, but in recent years precise photometry has yielded an abundance of accurate rotational periods for solar-type stars, especially in younger clusters, and this has enabled a much better view of the phenomena taking place.

\subsection{Rotational Spindown and Gyrochronology}\label{spindown}

The decline of rotation with age for solar-type stars is well-established from observations of clusters and field stars.  Because of the feedback scenario just outlined, this spindown is thought to be deterministic, not probabilistic, and there is a convergence in rotation rates with time among coeval populations.  Still, the details of AM loss remain to be investigated and there are poorly-understood aspects of spindown that challenge our picture, particularly for PMS and ZAMS stars.  Spindown and AM evolution have been addressed in many reviews; recent ones include \citet{barnes09}, \citet{irwin09}, and  \citet{mamajek09b}.  

\begin{figure}[t]
\includegraphics[scale=0.60]{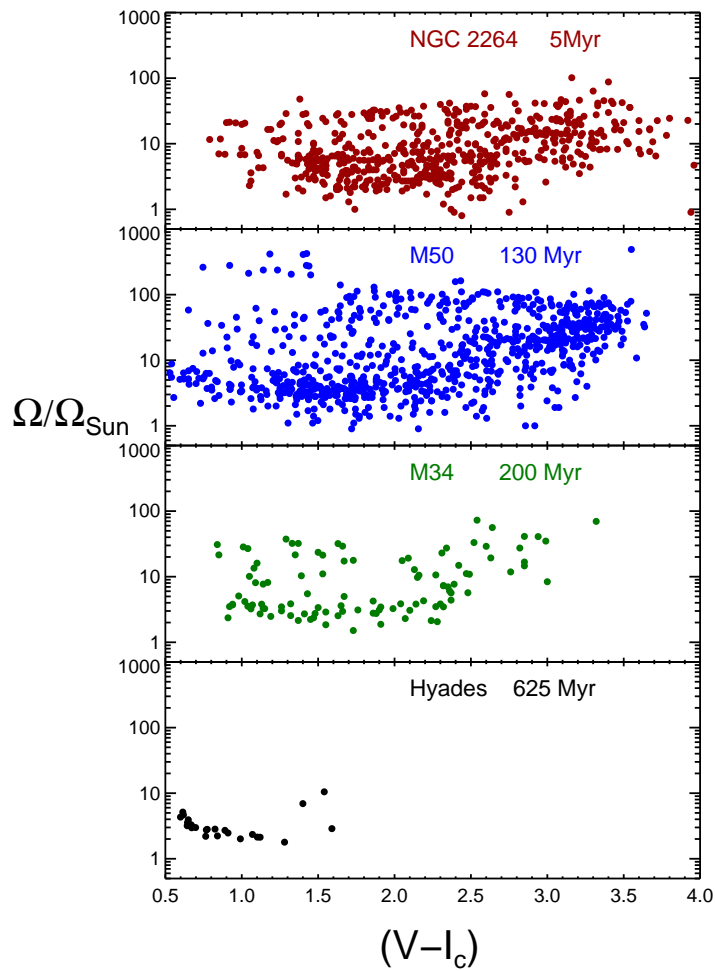}
\caption{\label{cluster-rotation}
Observations of \prot\ in young clusters.  The observed periods have been divided into the solar Carrington rotation (25.38 days) to yield an angular rate relative to the Sun.  Shown are data for NGC 2264 (age approximately 5 Myr), M50 (130 Myr); M34 (200 Myr); and  the Hyades to show how rotation has dropped and converged by the Hyades age of 625 Myr.  Observations of \prot\ are available for many other clusters (see \citet{irwin09}) but these are representative of the trends.  The data were provided by J. Irwin and the sources are listed in \citet{irwin09}.
}
\end{figure}

\citet{barnes07} has calibrated a relation between rotation period, color, and age, using data for clusters, and this was updated by \citet{mamahill08}.  These calibrations involve some assumptions that affect how they can be applied to individual stars.
\begin{enumerate}

\item As noted, OCs older than $\sim500$ Myr are relatively rare and so tend to be distant.  This inhibits testing the models in the realm they are most needed: among the older stars where one wishes to estimate ages of individual field stars.  This can be addressed to some extent by finding nearby binaries with intermediate-mass primaries and solar-type secondaries so that the system's age can be established from isochrone placement, but there are not very many such systems to work with.

\item \citet{barnes07} characterizes each cluster with a mean relation fitted through the stars with the longest periods.  Indeed, most members in young clusters are found on his ``I'' sequence, but in any young cluster there is a substantial fraction of more rapidly-rotating stars (his ``C'' sequence), in some cases spinning at 100 times the rate of the slower stars (Fig. \ref{cluster-rotation}).  These ultra-fast rotators (UFRs) are themselves a fascinating phenomenon with important implications for AM partition and evolution among PMS stars.  They are bona fide cluster members and add substantial cosmic ``noise'' and uncertainty to the calibration.  This can be seen clearly in \citet{irwin09}, who illustrate the available cluster data.
\marginpar{\bf Fig. \ref{cluster-rotation}: Spread in \prot\ with mass for representative clusters.}

\item In addition to adding uncertainty to age determinations, the scatter seen in young clusters means that ages estimated from \prot\ for young stars will have a bias.  The gyrochronology calibrations of \citet{barnes07} and \citet{mamahill08} are based on means for the slower rotators in the various clusters.  Those same relations, if applied to one of those clusters, would produce a distribution of ages that is asymmetric with a pronounced tail toward stars that appear to be younger than most of the cluster; these are the UFRs.  For an individual star one does not know where in such a distribution the star's AM lies.

\item This spread in AM appears to have converged by an age of $\sim500$ Myr, as shown by observations of rotation periods in F and G stars in the Hyades and Coma Berenices clusters \citep{collier09}.

\item Rotation rates at a given age are highly mass-sensitive, particularly above 1 \MS\ on the main sequence (PMS rates tend to be more independent of mass; see NGC 2264 in Fig. \ref{cluster-rotation}).  Calibrations are generally done in terms of stellar color as a proxy for mass, but the color-mass relation is different for different compositions and this adds systematic uncertainty.

\item Obtaining the \prot\ distribution in clusters is relatively straightforward in that young stars tend to show significant photometric modulation from surface inhomogeneities (spots) and clusters cover modest regions of the sky, making it possible to get the needed observations over a month or so.  By comparison, the older stars of the field must be observed individually and the amplitude of variation tends to be much less.  Rotation periods in young stars range from a few hours to $\sim2$ weeks, while older stars like the Sun have periods of a month or more.

\item Our knowledge of rotation from \prot\ is also biased because only some stars show periodic variability at a given time.  They are generally the more rapidly rotating and more active stars, but even among those not all stars show rotational modulation all the time.  A higher-contrast signal (compared to broad-band measures) can be seen by using observations of a chromospheric feature such as Ca {\sc ii} H and K, although at the cost of signal-to-noise.  Even then luck plays a role, as illustrated by observations of Mg {\sc ii} h and k in the Sun with the SORCE mission: For the most part the rotational modulation of the hk emission was unmistakable, but at times little or no modulation could be seen \citep{snow05}.

\item Rotation periods, when available, are clearly to be preferred as a precise and aspect-independent measure of rotation, but use of \vsini\ data is reasonable as long as rotation dominates the line broadening.  In such cases the errors are modest, and the statistics of rotation is such that most stars show a \vsini\ that is at or near the true equatorial velocity.  For example, the distribution of \vsini\ in the Pleiades \citep{rot5}, when converted to \prot, is just like that from \prot\ for the Pleiades or for other clusters of the same age \citep{irwin09}.

\item As with any astronomical phenomenon, we can only observe the OCs that nature makes available to us.  The available data is impressive and growing \citep{irwin09}, yet it is still difficult to assess how typical or not an individual cluster may be.  An even broader sampling of the nearby clusters is needed.

\item Using rotation to derive an age rests on the critical assumption that AM loss and rotational convergence are inevitable and that assumption remains untested.  It is called into question by the recent finding of a star in M67 (age 4 Gyr) that rotates at about twice the rate of other stars of the same color, which rotate at the solar rate, as expected \citep{reiners09}.  This observation needs to be confirmed (the object could be a binary, for instance), but an unambiguous spread in rotation at such a late age would ruin the carefully constructed picture we have created.  

\end{enumerate}

The rotation-age calibrations of \citet{barnes07} and \citet{mamahill08} distill knowledge from clusters into mean relations.  It is, of course, helpful to take advantage of data for many stars in a cluster to define a mean relation, but the scatter seen indicates how much different in age a star could seem to be without our realizing it.  The scatter seen is not statistical and not an error; it is real and potentially confusing.  \citet{collier09} show an example of how much spread in apparent age can exist in a coeval population: For the well-behaved Hyades and Coma clusters it is $\pm40$\%.  In other words, there is an inherent liability in applying knowledge of a group phenomenon to individuals.  At the same time, it should be possible to use these calibrations to obtain the age of a young ensemble of, say, 10 or more stars.

\subsubsection{SUMMARY OF GYROCHRONOLOGY}

\begin{description}
\item[$+$] Rotation periods can generally be measured with good precision and accuracy.  Also, rotation is not subject to variability except for the relatively minor effect of differential rotation.
\item[$+$] The dominant fraction of post-ZAMS and older stars ($\tau > 500$ Myr) are reasonably well behaved and exhibit a well-defined trend of rotation with age.
\item[$+$] A plausible and partially-tested scenario exists to explain the observed spindown and convergence in rotation with age.
\item[$-$] No age can be estimated if the rotation period is not detected, and that limits gyrochronology to solar-type stars (because they have spots) and favors younger stars for the same reason.  Late-F and early-G dwarfs tend to exhibit less of a rotational signature as well; note the lack of stars at the bluer colors in Figure \ref{cluster-rotation}.
\item[$-$] Obtaining rotation periods is observationally intensive and is not always possible if the level of photometric variations is too low.  Low amplitude can occur at all levels of activity and can bias the results.
\item[$-$] In some cases an alias of the true rotation period may be measured instead of the true period.
\item[$-$] Differential rotation on the stellar surface can mean that the measured rotation period is not the equatorial value.  However, experience so far shows that the uncertainty this introduces is modest compared to other effects.
\item[$-$] Significant scatter in \prot\ exists among stars in a cluster which are of the same age, particularly for $\tau < 500$ Myr (Fig. \ref{cluster-rotation}).  \citet{barnes07} argues that the stars in clusters that are on his ``I'' sequence follow the $\tau^{-1/2}$ law that appears to hold for older objects, but it is not possible to tell if a given individual star meets that criterion or not.  Given the way in which the rotation-age relation is calibrated (see above), the resulting ages can have a bias because the rapid rotators that are present at young ages are predicted to be much younger than they really are.
\item[$-$] Different clusters of the same or similar ages do not show identical distributions of \prot\ with mass \citep{irwin09}, adding additional inherent uncertainty.  There may be other parameters (such as composition) that current calibrations do not take into account, or initial conditions may matter even after stars reach the ZAMS.
\end{description}

\subsubsection{RECOMMENDATION}

Despite these known and potential flaws, using \prot\ to estimate an age for a G or K dwarf is one of the better-calibrated and better-understood methods that is available.   Paradigms are our creations to guide our understanding -- they do not necessarily represent reality -- but the rotation-age paradigm has withstood well the tests of it that have been made, given the admonitions discussed above.  The calibration of \citet{mamahill08} for F, G, and K stars is to be preferred because it appears to avoid a systematic error that \citet{barnes07} included.

\subsection{Activity and Age}

As mentioned, \citet{skumanich72} first suggested that a $\tau^{-1/2}$ power law may relate both rotation and activity to age.  In particular, he used measures of Ca {\sc ii} H and K chromospheric activity because those data were available, but there are other indicators of activity that can be used as well.  The Skumanich relations also imply that activity is proportional to rotation and that is indeed observed but with two critical limitations.  First, the activity we see may have its origins in rotation, but the chain of processes that lead from spin to magnetic field generation to observed activity is a long one with poorly understood links.  One result of that chain is that activity can be highly variable for a given star.  Second, for reasons not well understood, activity saturates at high rotation rates.  This makes qualitative sense in that activity cannot remain linearly proportional to rotation indefinitely, but it adds additional uncertainty.

What is broadly called activity is observed in a number of forms.  All of it is non-thermal emission from the star and is much weaker in flux than the photosphere.  As a result, activity is best observed at wavelengths where the photosphere is weak or suppressed, such as in x-rays, the ultraviolet, the cores of strong absorption lines, and in the radio regime.  For some of these wavelengths (x-ray and radio) the signature of activity is unambiguous in that the thermal background is non-existent.  However, the most easily observed indicators of activity are the cores of strong lines in optical spectra and in those cases disentangling the flux due to activity from the background is not so clear-cut.  In addition, it is not always clear how to parameterize the activity, with some observers preferring the flux in, say, soft x-rays.  Here I work with normalized indices, in which the observed flux is divided the stellar bolometric flux to derive a flux ratio such as $R_X$ or $R_{\rm HK}$ because these indices appear to produce consistent results across a range of spectral types.

\subsubsection{Ca {\sc ii} CHROMOSPHERIC EMISSION}

By far the most frequently observed manifestation of activity is the chromospheric reversals seen in the cores of resonance lines in optical spectra, particularly Ca {\sc ii} H and K.   Here I will focus primarily on FGK stars but HK data exist also for M dwarfs.  The use of Ca H and K as an activity indicator for stars was pioneered by Olin Wilson, first with photographic spectra and then with a spectrophotometer.  \citet{vaughan80} carried Wilson's work forward with a dedicated instrument that has been seminal for understanding the Sun in the context of the stars and vice versa.  The HK project at Mt. Wilson  has led to long-term monitoring of stellar activity cycles for nearly a half-century and has included a volume-limited survey as well as observations of a few clusters accessible with their telescope (the Hyades primarily).  Many papers and reviews have described the HK observations and their reduction to a physically relevant quantity, $R^\prime_{\rm HK}$, the ratio of the HK flux to \Lbol, corrected for photospheric light in the bandpasses (hence the prime; see \citet{noyes84}).   \citet{baliunas95} provide a summary of the Mt. Wilson program.

\begin{figure}[t]
\includegraphics[scale=0.80]{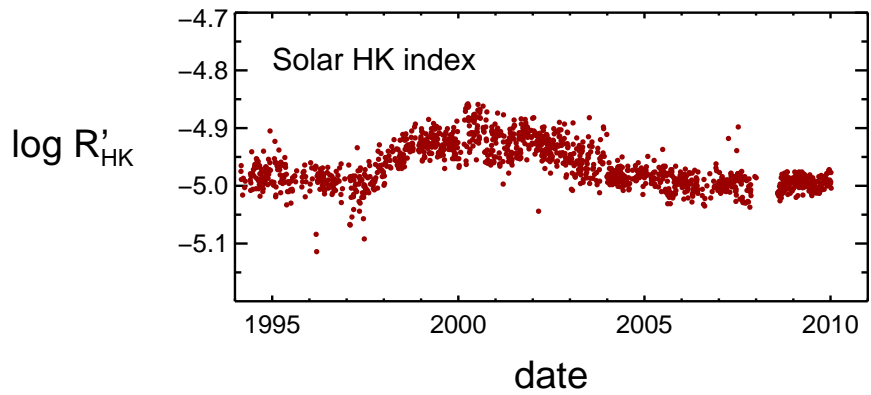}
\caption{\label{HK-Sun}
Observations of the disk-integrated core of the Ca II K line in the Sun.  The observations are from a Lowell Observatory program that monitors the activity cycles of solar-type stars (see \citet{hall07} for a description).  The solar observations are fed into the same spectrograph used for the stellar work using an optical fiber.  The time span of these data exceeds a solar cycle and the rise and fall of he overall activity of the Sun can be seen, together with variations.  The scatter is entirely astrophysical, not noise.
}
\end{figure}

Observing HK in FGK stars has a very significant advantage over attempting to detect \prot: A single low-resolution ($R \approx 2000$) spectrum suffices for estimating the mean activity level.  In addition, HK emission can be detected even in older stars that lack enough contrast to make their rotation detectable from modulation of surface inhomogeneities.  That makes it possible to measure HK in thousands of field stars.  However, HK activity is known to vary on a number of time-scales, from hours (flares), to days and weeks (rotation), to years (long-term activity cycles).  Each of these regimes is itself a subject of study for understanding stellar physics and the dynamo mechanism, but here our concern is that variability adds uncertainty in defining what the mean activity level of a given star is.  Still, these variations are of fairly low amplitude compared to the overall drop of activity with age.  Data for stars observed many times at Mt. Wilson show that one or two HK measurements produce an HK equivalent width that is within 8\% of the long-term mean \citep{soderblom85}.  This leads to about a 0.08 dex change in $\log R^\prime_{\rm HK}$  for an older solar-type star and less for cooler or younger stars.  \citet{wright04} compared Mt. Wilson observations to a similar index derived from high-resolution spectra for the same stars and saw a 13\% scatter.  Multiple, extended measurements of chromospherically-quiet stars show a scatter of about 6\% \citep{wright04}; this represents the floor to inherent variation in the best cases.  As another example, Figure \ref{HK-Sun} shows long-term observations of HK in the Sun \citep{hall07} obtained at Lowell as part of a stellar monitoring program.  Short-term variations can be substantial, and the solar cycle is easily seen.
\marginpar{\bf Fig. \ref{HK-Sun}: HK measures for the Sun at different times over a period of years.}

Following the initial look at the age dependence of activity by \citet{skumanich72}, \citet{SDJ91} undertook a more thorough calibration, confirming that the $\tau^{-1/2}$ relation appeared to work well for clusters and binary companions of more massive stars, with the saturation turn-over at young ages noted above.  \citet{SDJ91} also applied an additional method that took advantage of the unbiased (essentially volume-limited) nature of the sample that was available.  That method used the Sun to anchor the relation at 4.5 Gyr and assumed HK emission declines monotonically with age (variability is not critical as long as the age trend dominates) and that the Galactic star formation rate in the vicinity of the Sun has been constant.  The result, after correcting for ``disk heating'' in the Galaxy \citep{wielen77}, was essentially the same as the directly-calibrated relation, suggesting both methods are reasonably sound and that the chromospheric data are consistent with a constant star formation rate.

\begin{figure}[t]
\includegraphics[scale=0.60]{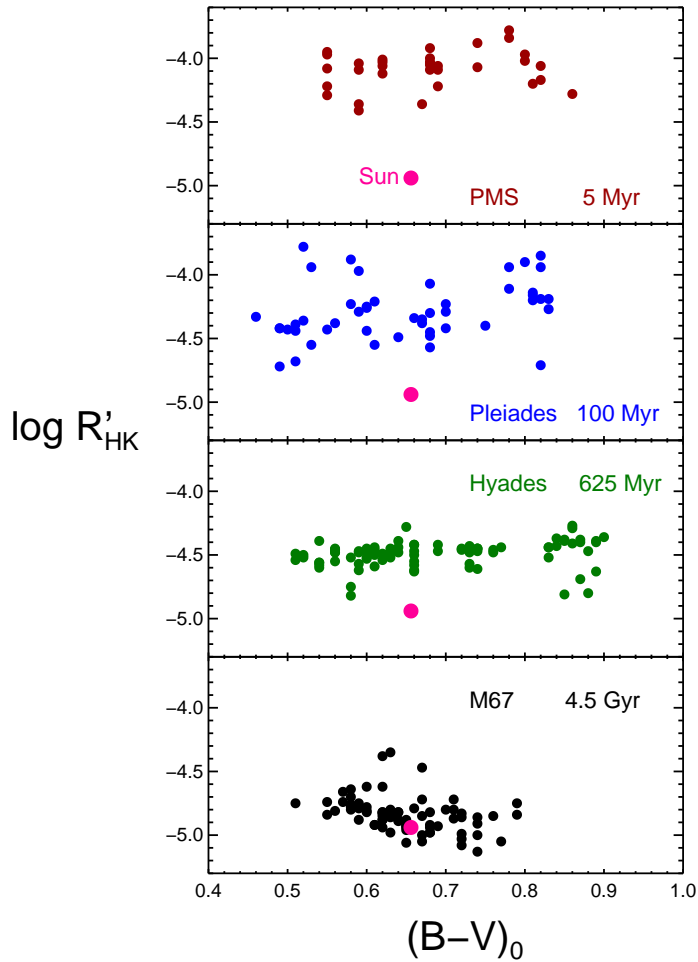}
\caption{\label{HK-spread}
Observations of Ca {\sc ii} H and K activity at representative ages (adapted from \citet{mamahill08}).  The ordinate is a normalized index of HK strength, described in the text.  Shown are observations for PMS groups (red; age 5 Myr); the Pleiades (blue; 100 Myr); the Hyades (green; 625 Myr); and M67 (black; 4 Gyr).  The solar point is shown as a pink dot.  Note in particular the range of HK indices seen in any one cluster and that the spread within the Pleiades, for example, exceeds the net change in the HK index in going from the mean Pleiades value to that of the Sun.
}
\end{figure}

Just as for rotation, HK measures in young ($<500$ Myr) clusters show lots of scatter that is inherent (Fig. \ref{HK-spread}).  
\marginpar{\bf Fig. \ref{HK-spread}: HK in clusters.}
This scatter is related to the spread in rotation and it exceeds the differences in cluster averages so that there is significant overlap in apparent age for stars with the same HK level.  As with rotation, this scatter is largely absent in the Hyades and it is assumed to reflect the convergence in rotation that has taken place.  For older stars, the HK emission gets weak and it is not clear which portion of the line to attribute to just the chromosphere.  

Those early analyses assumed all the stars were of solar composition and that there was no dependence of the HK emission on composition.  That assumption is certainly not true for the oldest stars in the solar neighborhood.  Lower metallicity may affect the dynamo itself by changing a star's structure, but, at a minimum, decreasing [Fe/H] in a star means decreasing [Ca/H] and so changing the line formation conditions.  The net result is to make a metal-poor star look more active and younger than it really is because the line cores are not as deep.  \citet{RPM98} and \citet{RPM00} have calibrated this effect.  They also showed that high levels of activity may affect Str\"{o}mgren photometry of stars and thus ages determined in that fashion as well.

In addition to using HK to estimate ages, an essential question has been the overall form of the relation and its implications both for ages and for stellar physics.  A $\tau^{-1/2}$ power law may be mathematically simple yet physically incorrect, and the question arises in part because the distribution of HK strengths for solar-type stars shows a ``Vaughan-Preston gap'' that corresponds to an age of $\sim$1-2 Gyr.  \citet{pace09} have addressed this issue by obtaining HK observations of some members of clusters in that age range.  \citet{pace04} and \citet{pace09} suggest that HK emission ceases to drop after $\sim1.5$ Gyr because of their observations of HK in several clusters near 1-2 Gyr age.  Verifying and calibrating the continuing decline of activity in older stars is difficult, both because of the scarcity of old clusters and because the chromospheric emission is weak and so high signal-to-noise spectra are required.  However, both the Sun and the $\alpha$ Centauri system, at 4-5 Gyr, show weaker HK emission than 1 Gyr old clusters, and M67 -- also about 4 Gyr old -- also shows weaker HK emission (Fig. \ref{HK-spread}) but also a spread in HK that exceeds observational error \citep{giampapa06}.  There is still much we do not know about the evolution of rotation and activity in stars like the Sun.

\subsubsection{OTHER CHROMOSPHERIC INDICATORS}

Other useful measures of chromospheric activity in late-type stars include H$\alpha$ and Mg {\sc ii} h and k.  \citet{lyra05} present observations and an analysis of the age dependence of H$\alpha$ in solar-type stars, an effort started by \citet{herbig85}.  H$\alpha$ is not inherently as deep a feature as the H and K lines and so the chromospheric component is not as easily detected, requiring both good resolution and high signal-to-noise.  The Mg {\sc ii} h and k lines are very similar to the Ca {\sc ii} H and K lines and formed under similar conditions.  \citet{cardini07} show a study of Mg {\sc ii} hk emission versus age.

Both H$\alpha$ and Mg {\sc ii} hk lack the large datasets that are available for Ca {\sc ii} HK.  However, in some cases those indices may be available for a star of interest when HK is lacking.  Given that many fewer cluster observations are available to calibrate H$\alpha$ and Mg {\sc ii} as a function of age, the best use of them may be to first transform them into an effective HK index using mean relations and then use the HK versus age calibration.

\subsubsection{CORONAL EMISSION}

The coronae of stars like the Sun reveal themselves in soft x-rays.  This can provide a high-contrast signal because stellar emission at those wavelengths is entirely non-thermal, yet the signal is a weak one and difficult to detect except for nearby and young stars.  The significant strength of x-ray emission in PMS and ZAMS late-type stars has made survey missions such as {\it Rosat} an effective means of identifying previously unknown very young stars in the field \citep{Zuck04}.

The limited data available means that there is not an independent calibration of x-ray emission versus age, 
although \citet{gudel97} studied a small sample of solar-type stars with ages determined in other ways and found $L_X \propto \tau^{-1.5}$, a much steeper age dependence than the $\tau^{-1/2}$ seen for HK emission.  The situation for x-rays also involves substantial changes in coronal structure with time as the temperatures change, and so the measured decline depends on the x-ray wavelengths included in the observation.  It has also been shown \citep{sterzik97} that x-ray emission and Ca HK emission are realted to one another with a power law over nearly four decades of x-ray flux, with a scatter of about 0.06 dex in $R^\prime_{\rm HK}$.  \citet{mamahill08} show a mean relation to convert x-ray flux into HK.  There is no advantage to using x-ray emission as an age indicator if HK data are available, and, in addition, the x-ray fluxes are significantly more variable that is HK, adding uncertainty.  Like chromospheric emission, x-ray emission saturates at high levels \citep{pizzo03}.

\subsubsection{SUMMARY OF USING ACTIVITY TO ESTIMATE AGE}

\begin{description}
\item[$+$] Obtaining an activity index for a star is observationally straightforward and so large datasets exist.  More activity data exist than for any other age indicator, particularly for field stars.
\item[$+$] Ca {\sc ii} H and K are easily observed from the ground and only moderate resolution is needed to get a measure of the activity.
\item[$+$] Although activity is variable on nearly all time-scales, for main sequence stars that variability is generally much less than the broad decline of activity with age and so even a single observation provides a reasonable estimate of the mean level.  Ages can be estimated to about 0.2 dex \citep{SDJ91}.
\item[$+$] X-ray emission can also be used.  It is generally detectable only for young stars, but it presents a high-contrast signature.
\item[$+$] For a given age, activity is only moderately mass-dependent \citep{mamahill08}.
\item[$-$] Activity arises from the interaction of convection and rotation (as far as we know) and so it follows from the rotation, but with significant variability added.  In other words, activity is secondary to rotation, and a given rotation rate in a star can result in a range of activity.  This paradigm needs further study and verification, and could be wrong in important respects.  Also, activity in these forms is only seen in convective stars, about F6V and later.
\item[$-$] Young (PMS and ZAMS) stars exhibit a broad range of activity at a given age.  This prevents activity from being used to date single stars, and ensembles of at least 10-20 stars are needed to define a mean relation.
\item[$-$] Activity can be difficult to detect in older stars, and the portion of the signal to attribute to age (as opposed to a basal level expected to be present in a minimally active star) is uncertain.  We know the Sun has experienced periods such as the Maunder minimum in which spots are rare, but we do not know the activity level one would observe in those cases.
\item[$-$] Currently-measured activity indices (especially Ca {\sc ii} H and K) probably contain a sensitivity to metallicity that has not been fully taken into account.  One result is that metal-poor stars can appear younger than they really are.
\item[$-$] Activity indices such as $R^\prime_{\rm HK}$ are generally measured in a way that does not take into account rotational broadening of lines, leading to systematic misestimates.  \citet{schroeder09} show how to account for this effect.
\item[$-$] Activity is inherently variable on all time-scales, and so extended measurements are needed to get the best estimate of a mean level.
\item[$-$] Activity saturates at high rotation rates so that activity becomes only weakly dependent on \prot\ and so is a poor age indicator in that regime.
\item[$-$] Comprehensive data are available only for Ca {\sc ii} H and K.  Other activity measures have not been calibrated against age.
\end{description}

\subsubsection{RECOMMENDATION}

Activity observations need to be first reduced to an index such as $R^\prime_{\rm HK}$ that is normalized to the star's luminosity (see \citet{wright04} for the procedure for Ca HK).  There are extensive observations for Ca {\sc ii} H and K, including stars in clusters, but many fewer data for other activity indices as well as fewer observations of clusters.  That makes it advantageous to first convert an activity observation into the HK equivalent with a mean relation such as that shown in Figure \ref{Lx-HK}.

Many HK-age relations have been published.  In \citet{SDJ91} an activity-age relation was presented which was calculated for a constant star formation rate that incorporated a correction for disk heating.  Because this was not analytical, \citet{SDJ91} did not provide a formula to go with that curve (shown in their Fig. 10), but the following relation is a close approximation for $\log R^\prime_{\rm HK}$ values from $-5.1$ to $-4.3$:
\begin{equation}
\label{eqHK}
\log \tau_{\rm HK} = -8568.2 - 7037.6x - 2164.6x^2 - 295.76x^3 - 15.144x^4,
\end{equation}
where $x = \log R^\prime_{\rm HK}$ and $\tau_{\rm HK}$ is in yr.  I emphasize that this formula is a convenient mathematical fit to the means for clusters and to the Sun; it is not a physical relationship and it does not take into account the considerable spread seen at any one age.  \citet{wright04} present a similar polynomial; the differences between the two are $<5$\%\ above 2 Gyr, but the \citet{wright04} ages are systematically greater than \citet{SDJ91} below that.  The turnover shown in \citet{SDJ91} corrects for this effect, which is due to saturation of the HK emission.

\subsection{Lithium Depletion}\label{Li-depletion}

\begin{figure}[t]
\includegraphics[scale=0.60]{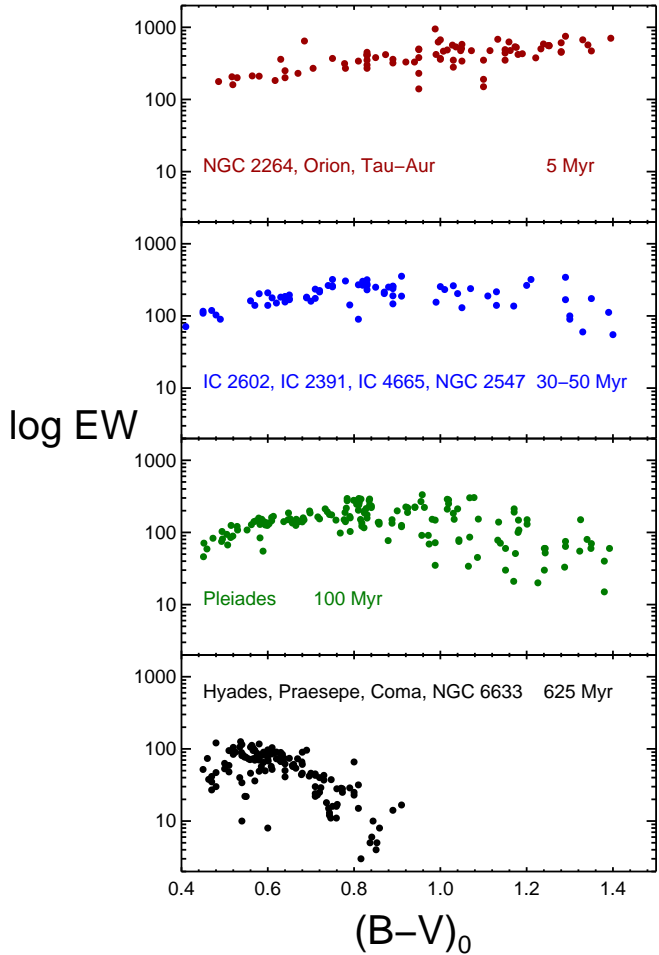}
\caption{\label{Li-spread}
Observations of lithium in clusters, using data from  from \citet{sestito05}.  Shown are data for T Tauri stars at about 5 Myr (red), clusters from 30 to 50 Myr old (blue), the Pleiades at 100 Myr (green), and clusters at about 600 Myr (black).  The data are shown in observational units to illustrate the inherent scatter seen.  This scatter is real and not error.  Note that the T Tauri stars are undepleted in Li and exhibit little scatter in EW, while stars in the Pleiades have a spread in EW of 1 dex for \bvz $>1.0$ (about K0V and later); the spread in abundance is even larger.
}
\end{figure}

We know that stars like the Sun deplete their surface Li abundances over time because primordial solar system material has about 200 times the Li content of the present solar surface.   Just as for rotation and activity, young OCs show that young solar-type stars have much more Li than MS stars and also that there is substantial and real scatter in Li at those young ages (Fig. \ref{Li-spread}).  Some of that scatter may correlate with rotation \citep{soderblom93a}.
\marginpar{\bf Fig. \ref{Li-spread}: Li in clusters to show scatter.}

Li is observed in stars as absorption at the 6708 \AA\ resonance doublet.  This feature is analogous to the Na D lines or Ca {\sc ii} H and K but is unresolved.  The weakness of this feature in the Sun ($W_\lambda < 2$ m\AA) demonstrates the element's paucity in the solar atmosphere.  The 6708 \AA\ feature is due to Li {\sc i}.  The ionization potential of Li is low and so most of it is in the form of Li {\sc ii} which is invisible.  Because of this, the formation of the 6708 \AA\ feature is highly sensitive to \teff.  Non-LTE effects can also be significant, particularly for Li-rich stars \citep{lind09}.  

Detecting the Li 6708 \AA\ feature in young stars is easy ($W_\lambda > 100$ m\AA) and indeed the presence of the feature is used as a discriminant for T Tauri stars.  A secondary Li feature is sometimes seen in Li-rich stars at 6103 \AA.  Detecting the Li feature in Li-poor stars like the Sun is very difficult because the 6708 \AA\ feature is blended with many other lines of comparable strength.  Extracting the solar Li abundance was a tour de force, both in the observations \citep{brault75} and in the analysis \citep{muller75}.

\citet{sestito05} summarize and analyze the available Li data for OC stars.  These data are consistent with all the clusters have formed with a uniform initial Li abundance of about 3.2 (on a logarithmic scale where H = 12).  They show that most Li depletion occurs before stars reach the ZAMS, with some continuing depletion up to at most 1-2 Gyr.  The OC data also show that Li depletion is not a continuous process and cannot be described by a simple $\tau^{-\alpha}$ type of law.

None of the models proposed for Li depletion can fully explain the observations, but there are several factors that may contribute  to fundamental uncertainty in interpreting Li abundances.  First, because of the temperature sensitivity just noted, the observed $W_\lambda$ could be altered by stellar surface inhomogeneities (spots).  In other words, starspots could lead to an apparent spread in Li where none really exists.  \citet{soderblom93a} show that the degree of spottedness would have to be extreme to produce the scatter in Li seen in the Pleiades, for instance.  Also, T Tauri stars, which we observe to have very high levels of spottedness, show less Li scatter than ZAMS clusters do \citep{sestito05}.  \citet{soderblom93a} looked for Li variability that might arise from spots and saw none, but such an effect has not been fully ruled out.

A second potential effect that could add scatter would be late deposition of gas-depleted material onto a star's surface after planets form.  The apparent Li abundance would be especially enhanced because the star's surface Li content is low to begin with.  This process would be expected to lead to higher Li enhancements in, say, a late-F dwarf compared to a K dwarf because the F star has much less mass in its convective envelope.  Searches for this effect have seen none and there is no conclusive evidence that stellar Li abundances are related to the presence of planets.

\citet{mentuch08} used observations of Li in young, nearby associations and fitted them to calculated Li isochrones to derive ages.  This has the advantage of using data from a number of stars to arrive at a net age but makes the results model-dependent.  \citet{nasci09} go much further and use models of Li depletion with non-standard mixing processes to derive ages for stars as old as the Sun.  Doing so requires assumptions about the histories of the stars that cannot be tested, particularly their angular momentum evolution.

The depletion of Li in low-mass stars is a significant subject in itself and cannot be fully covered here.  The mechanism is believed to be convection (or convective overshoot) combined with non-standard mixing and observations of Li are important for probing fundamental aspects of solar-type stars.  However, the use of Li for estimating ages is problematic for a number of reasons in addition to those just noted.  Late-F and early-G dwarfs deplete little Li and the scatter seen within clusters exceeds any age trend.  Li may work better for late-G and K dwarfs, provided that a group of 10-20 stars is examined to average out the inherent scatter.

\subsubsection{SUMMARY OF LITHIUM DEPLETION}

\begin{description}

\item[$+$] The primary Li feature at 6708 \AA\ is easy to detect and measure in young solar-type stars, having an equivalent width of 100 m\AA\ or more and lying near the peak of sensitivity for many modern spectrographs.  Good resolution (at least $R \approx 20,000$) and good signal-to-noise are needed, however.

\item[$+$] The trend of declining Li abundance with age is strong for PMS stars, particularly K dwarfs.

\item[$+$] Many clusters have been observed to provide a calibration.

\item[$+$] Clusters appear to form with a uniform initial Li abundance.

\item[$+$] The youngest stars (T Tauris) all appear to be rich in Li, confirming our expectation.  There may be exceptions \citep{white05}, but in such small numbers that special circumstances may apply.

\item[$+$] Young FGK stars appear to always show a strong Li feature, making its presence a discriminator of youth at least, even if a quantitative age is difficult to establish.

\item[$-$] Little change is seen in Li with age for late-F and early-G stars.

\item[$-$] The physical processes that drive Li depletion remain poorly understood.  Models can reproduce the solar abundance \citep{charbonnel05} but detecting Li in stars as old or older than the Sun is very difficult and so there are few constraints on the theory at present.

\item[$-$] PMS and ZAMS stars of a common age exhibit a large range of apparent Li abundance for reasons that are not understood.  As is the case for rotation and activity, a single star can be placed among clusters with a broad range of age, and so an ensemble of 10-20 stars is needed to define a mean relation.

\item[$-$] Converting an observed Li equivalent width to an abundance is highly temperature sensitive and can require a significant non-LTE correction.  Despite this, one can work with just the equivalent width and color to compare an ensemble of stars to clusters.

\item[$-$] The apparent Li abundance of a star may reflect unappreciated processes that could add Li-rich material to the surface (left over from planet formation) or could remove the star's surface layers during an early phase of high mass loss \citep{sackmann03}.  Starspots may alter the equivalent width of the Li feature for a given abundance.  PMS magnetic fields may inhibit Li depletion \citep{ventura98}.

\end{description}

\subsection{Photometric Variability}

All stars vary to some extent.  Stars like the Sun are known to vary on many time-scales, and the dominant cause is surface inhomogeneities (spots).  Younger stars rotate faster than older stars and have higher activity levels and greater spot coverage.  As a result, there is an overall trend of the degree of photometric variability with age.  This is not yet well enough characterized to be useful as an age indicator, and, in any case, variability implies scatter and uncertainty.


\section{STATISTICAL METHODS}\label{statistical}

There are two broad Galactic trends that relate to stellar ages: the age-metallicity relation (AMR) and the increase in disk scale height with age as stars and clusters encounter massive objects in the Galaxy.  The AMR will not be treated here because its existence is in question but also because it is not very useful.  For example, among stars formed in the current epoch there is at least a factor of two range in metallicity.  \citet{wheeler89} have suggested that [O/H] may be a  better chronometer than [Fe/H] and that may be useful for populations but not for individual stars.  See \citet{feltzing09} for more on the AMR.

The other phenomenon -- known as disk heating -- is generally studied as a means to understanding some Galactic dynamical processes \citep{wielen77}.  It can also be inverted to determine a mean age for a population if one has full, three-dimensional kinematic data.  First, disk heating is an incompletely understood process and it is not clear what objects -- giant molecular clouds, perhaps, or massive black holes -- produce the observed effect.  In any case, what is observed is a monotonic increase in the scale height of populations in the Galaxy perpendicular to the Galactic plane as one looks at progressively older ensembles of stars.  Second, one needs full, three-dimensional kinematic information on an unbiased sample, preferably volume-limited.  That means having good parallaxes and radial velocities in addition to proper motions.

The classic reference on the subject is \citet{wielen77}, who used Cepheids, nearby main sequence stars, and K and M dwarfs.  The latter sample was divided by HK emission class, a rough age indicator available at the time.  The main sequence stars were assigned ages to be half the MS lifetime for the mean of the sample.  (This means that he did not correct the ages for the phenomenon that he was reporting on.)   \citet{wielen77} used the KM dwarfs because there were sufficient numbers of them close enough to the Sun to have good parallaxes at the time (before {\it Hipparcos}).  

As an example, \citet{soderblom90} used kinematic data to show that the BY Draconis binaries have a mean age of 1-2 Gyr and that the lack of old BY Dra systems was probably due to coalescence of the binary systems as they lost AM.  \citet{reiners2009} applied the method to a volume-limited sample of late-M dwarfs to derive a kinematic age of 3 Gyr.  This is the same as the median age of G dwarfs in the solar neighborhood \citep{SDJ91} and results from the samples telling the same story: both G- and late-M dwarfs represent the full age range of the Galactic disk ($\sim10$ Gyr), and, given disk heating, the removal of the oldest stars from the solar neighborhood brings down the median age.

The paper by \citet{reiners2009} shows well how to apply the kinematic method to a sample.


\section{THE LIFE CYCLES OF AGES}

Each phase in a star's life offers specific problems and opportunities for estimating ages.  

\subsection{Newly-Formed Stars}

The time-scales involved in the earliest phases of a star's life are 0.1 to 1 Myr or even less and so are very challenging to address.  To a significant extent our assignment of the sequence of PMS phases is based on our preconceptions of how these systems ought to evolve, not actual measurements.  The co-existence of several of these phases within a given star-forming region is evidence that our understanding is incomplete.  \citet{evans09} present a comprehensive look at the formation phases in nearby star-forming regions and estimate relative time-scales and durations, and \citet{wyatt08} reviews the evolution of debris disks.

Also pertinent for this age range is an alternate model for early accretion that may affect ages at the $\sim0.1$ Myr level \citep{wuchterl03}.

\subsection{The Pre-Main Sequence}

Many of the most critical questions we ask about the formation and evolution of stars and planetary systems have to do with their lives before they reach the ZAMS, yet this phase presents severe challenges for determining ages.  The most-used method is isochrone placement (Sec. \ref{PMS}), which is itself limited in every single aspect of its application, from knowledge of \teff\ and \Lbol\ to the inadequacies of the models.  \citet{hillenbrand09} provides a review of age-dating techniques for PMS stars using isochrones and other methods.  Some other approaches are provided by \citet{mayne07}, who use empirical PMS isochrones to place clusters in an age order based on spline fits in magnitude versus color coordinates.  This is a sensible means of applying the most basic information available, although the age ordering could be affected by differing age spreads or differing binary frequencies or systematically erroneous distances, among other effects.  \citet{mayne07} also try to place a ``radiative-convective gap'' in the CMD; this feature makes sense in terms of the physics of models but in practice appears problematic to locate precisely.  Finally, \citet{naylor06} and \citet{naylor09} describe a method for fitting the MSTO that takes account of evolution among massive stars.  Their results yield ages that are 1.5 to 2 times those derived from PMS isochrones for very young clusters ($\sim 5$ Myr old).  This method is applied objectively but has the basic weakness of fitting the MSTO in a young cluster: a few of the most massive stars dominate the solution and their properties have the same vulnerabilities as other young stars (extinction, duplicity, etc.).

Although it can only be applied to a group of stars, the detection of the lithium depletion boundary (LDB) is an especially promising technique for PMS ensembles in theoretical terms because the models used are straightforward.  In practice the LDB method is of very limited use because it depends on obtaining good spectra of extremely faint objects, and that has been possible so far for only five nearby clusters (see Sec. \ref{LDB}).

Asteroseismology may provide some help in dating PMS stars but will be limited because our models of PMS stars are inadequate and because detection of oscillations in the presence of photometric variability is problematic.  At present the relative faintness of most PMS stars puts them beyond the reach of the available seismic methods.  However, more massive PMS stars lie in or near the instability strip and can show detectable oscillations \citep{zwintz08, guenther09}; these have been used more to test models of these stars than to establish ages.  BDs in very young associations should also lie near the instability strip \citep{cody09} and may reveal oscillations as well.

The empirical methods should in principle work well for PMS stars because the changes in rotation, activity, and lithium abundances that occur in the PMS phase are large and rapid.  But, at the same time, coeval populations of PMS and ZAMS stars in well-studied clusters show spreads in these quantities that exceed the slopes of the relations and there are saturation effects that lead to ceilings in these quantities for many stars.  In other words, it is possible to apply the empirical methods if one has a group of 10-20 stars or more so that a mean relation can be determined, but the methods cannot be used reliably on individual stars.

The movement of a PMS star in an HRD slows as it nears the MS, being roughly linear in terms of $\log \tau$.  Because of this there should be many more almost-on-the-ZAMS stars than there are T Tauri stars; these are the ``post-T Tauris.''  The practical difficulty is that the post-T Tauris have had enough time to drift out of their formation regions into the field and thus are not so easily found.  The major advance in this area in recent years has been the recognition that there are groups of PMS stars in the solar neighborhood that can be identified on the basis of their activity and space motions \citep{Zuck04}.  Just as for the star-forming regions, these post-T Tauri ensembles often contain objects in a range of physical states, from classical T Tauri stars with active accretion, to``naked'' T Tauris and post-T Tauris, emphasizing our poor overall understanding of these phases of stellar life.  The TW Hydrae Association is a particular case in point.  The ages of these ensembles have been estimated from dynamical expansion \citep{mamajek05} and lithium abundances \citep{mentuch08}.  In all cases it is the average age of the ensemble that is determined.

\subsection{Zero-Age Main Sequence and Near-ZAMS}

As is well known, stars change very little in their structure for the first $\sim1/3$ of their MS lifetimes; this is why they form a well-defined {\em main sequence}, after all.  For massive stars, their MS lifetimes are brief and, in part because of that, they are almost always found in star-forming regions and associations.  In that case the ages can be estimated either from the MSTO of more massive stars, or from where less-massive stars are just reaching the ZAMS.  The practical problems in these cases are those of open clusters (see Sec. \ref{OC-age}).  Isolated massive stars are rare but do exist; estimating their ages is difficult because the luminosity is uncertain, but it is sometimes possible when a spectroscopic gravity indicates the evolutionary state.

Intermediate-mass stars (late B through mid-F) can be difficult to age-date if they are isolated and removed from a context provided by a cohort.  However, A and F stars in particular have had temperatures and gravities determined from Str\"{o}mgren photometry \citep{olsen94}.  These can be calibrated against nearby open clusters which span an age range comparable to the MS lifetimes of A and F stars, but the uncertainties are again significant at least in relative terms.  Recent advances in stellar interferometry, however, have shown that some A stars are measurably non-spherical, meaning that traditional one-dimensional models cannot predict their surface properties and that we may have little idea of their evolutionary state.  

Solar-type stars on and near the ZAMS are discussed elsewhere in this review.  Lower mass stars (late K and M), however, remain challenging.  The empirical methods that work for G dwarfs are less helpful at lower masses.  Getting an age from a rotational period is still effective, but lithium is depleted far too quickly to be of use in most cases.  Activity in the lowest-mass stars is notoriously variable, even for old stars, although \citet{west08} have studied activity as a function of age in M dwarfs by using a large sample and statistically deriving ages of subsets from their kinematics.

\subsection{Subgiants}

Subgiants are the ``sweet spot'' in the HRD for determining ages from isochrone placement.  These stars are clearly well-removed from the main sequence and are changing quickly in luminosity and \teff.  These ages are obviously model-dependent, and metallicities are needed to know which set of models to which to compare, and the fundamental limitations are in the models themselves and our knowledge of the physics of such objects.  Intermediate-mass stars and lower masses develop or deepen CZs, and this can alter surfaces abundances of some elements.  Str\"{o}mgren photometry can be used to determine \teff\ and $\log g$ for A, F, and early-G stars \citep{olsen94} to place them in HRDs, but relative uncertainties are significant.

\subsection{Very-Low-Mass Stars}

Solar-type stars age slowly but very-low-mass (VLM) stars evolve hardly at all once they reach the ZAMS.  However, like all stars, prior to the ZAMS they change significantly in \Lbol\ \citep{burrows01} as do brown dwarfs (BDs).  All of these objects that are near the bottom of the stellar mass function are of interest in part inherently, in part for their role in the Galactic mass budget, and in part because they form the transition from bona fide stars to planets.  The obvious way to limit the age of a VLM object is to associate it with another star or group for which the age is better constrained.  Lacking that, these objects are challenging.

\citet{jameson08} describe a method to determine ages of L dwarfs from near-infrared photometry and distances when those objects are younger than the Hyades (i.e., $<0.7$ Gyr).  The method uses $M_K$, the absolute $K$-band magnitude, and is calibrated against nearby clusters.  Their estimated uncertainty in $\log \tau$ is 0.2 dex.  \citet{burgasser09} were able to place a lower limit on the age of a binary system composed of an M dwarf with a T dwarf, again because having a binary constrains models significantly.  However, association of a VLM object with another group is not always the solution: \citet{bonnefoy09} show that a VLM binary thought to be a member of the $\sim10$ Myr-old TW Hya association is itself about 30 Myr old and so unlikely to be a bona fide member.

\citet{burgasser2009} lists five age estimation techniques for VLM objects and compares them.  \citet{dupuy09} illustrates several case studies applying these methods to brown dwarfs.  \citet{west06} studied H$\alpha$ activity in a large sample of Sloan Digital Sky Survey M7 dwarfs and were not able to delineate an activity-age relation but found those low-mass stars decline in H$\alpha$ rapidly at an age of 67 Gyr.

\subsection{Very Old Stars and Population II}

Most of the Galaxy's oldest stars are too distant to currently have good parallaxes which would allow the luminosity to be calculated to place the star in an HRD.  Instead it is possible to use nucleocosmochronometry (Sec. \ref{nucleo}).  Another possibility is using Be abundances \citep{smiljanic09}, similar to using Li for Pop I stars.


\section{CALIBRATING AGE INDICATORS}

The most fundamental problems in applying age estimators to stars arise from the difficulty and uncertainty of calibration.  Two methods will be discussed here: using clusters to define a property at specified ages, or taking advantage of having a volume-limited sample.  Both methods require assumptions to be made, and those assumptions can be difficult to test.

\subsection{The Dissolute Lives of Open Clusters: The Sun as Anchor}

The obvious way to calibrate an age relation is to observe stars in clusters.  This still leaves the problem of establishing the ages of clusters, but clusters have many stars and so it is usually possible to determine mean properties with good precision.  The underlying astrophysical limitation is that basic Galactic forces tear OCs apart on a time-scale of $\sim200$ Myr \citep{janes88, janes94} and more weakly-bound clusters can simply fall apart from their own internal motions more quickly than that.  
Old open clusters are inherently rare and not as near as younger examples.  Making matters worse, the age-related quantities to be measured (rotation, activity, and lithium in particular) in the older stars get progressively weaker, requiring ever higher signal-to-noise spectra and high resolution for ever fainter stars.

This process of cluster dissolution, of course, is where field stars come from, and so the stars we most want to place ages on are least represented among potential calibrators.  There is also a potential bias in calibrating against the few old clusters we can observe.  An old cluster such as M67 was probably born unusually rich and dense to have survived in the Galaxy for so long.  As such, its properties may not be fully representative of field stars of similar age, which came from more ordinary backgrounds.  In particular, the partitioning of angular momentum within a dense cluster could be very different from that for a loose association and that could have ramifications even late in life for stellar spindown.

On the other hand, among the older stars we have the Sun, for which many properties are precisely defined, and so the Sun is generally used to anchor age relations using its behavior at 4.5 Gyr.  This naturally raises the question of whether the Sun is itself representative of a 4.5 Gyr-old 1.0 \MS\ star.  \citet{gustaf08} (and others) have argued that the Sun is peculiar or unusual in one respect or another compared to other G dwarfs.  But given the many well-defined properties that characterize the Sun, it would be odd if several did not deviate significantly from statistical means, just as each of us is both typical and unusual as a human.  The null hypothesis should be that the Sun, on the whole, is a typical star for its mass, composition, and age, and that hypothesis has not been convincingly disproved.

\subsection{Clusters as Benchmarks}

Age determinations for OCs will be discussed in Section \ref{OC-age}.  They clearly provide the primary benchmarks we use to study age-related properties, not just to define a mean for a given age but also to examine how well a coeval population behaves.  Also, as noted in Sec. \ref{age-spread}, the available evidence suggests that any age spreads within clusters must be small, $\sim1$ Myr, and so the dramatic scatter we see in various properties among stars in ZAMS clusters must be explained in terms of the phenomenon itself.

\subsection{Volume-Limited Samples}

Volume-limited, or at least unbiased samples provide a special opportunity because all ages must be fairly represented.  That being the case, other criteria can be used to study ages and age relations, although assumptions must be made.  For example, \citet{SDJ91} studied a volume-limited sample of activity in G dwarfs and were able to derive an age-activity relation by using the Sun to anchor 4.5 Gyr and assuming that activity declines monotonically with age (or at least that the decline with age is dominant compared to variability), and then also applying a correction for ``disk heating'' to account for the lack of very old stars in the solar neighborhood.  \citet{reiners2009} have used a volume-limited sample of late-M dwarfs to determine a mean age by calculating their Galactic kinematics and comparing to observations of disk heating (see Sec. \ref{statistical}).

\subsection{The Special Role of Binaries}

Binary systems can be especially helpful in calibrating and checking age relations.  First, binary components provide a consistency check in that both stars should lie on a single isochrone (to within the errors) for a given age indicator if that indicator can be measured in both stars.  For example, \citet{barnes07} used rotation observations of well-separated visual binaries to confirm his gyrochronology calibration.  HK activity fares less well in the study of \citet{mamahill08} when components in binaries are compared, but most systems meet this expectation, even if there are conspicuous exceptions.

Second, binaries provide ``mini-clusters.''  A system with, say, a solar-type secondary and an intermediate-mass primary has its age limited by the HRD location of the more massive star.  This method was used by \citet{SDJ91} in calibrating HK emission versus age.  The obvious limitation is that a binary provides just a single datum, nowhere near the ensemble average one can achieve for a cluster.  At the same time, a system such as $\alpha$ Centauri provides not just well-studied stars but well-determined masses, and that is rare.

Binaries can also be helpful on setting limits to an age when one star is significantly more massive than the other.  \citet{gahm83} and \citet{lindroos83}, for example, sought F, G, and K dwarf companions to O and B stars as a means to identifying potential post-T Tauri stars, solar-mass stars that are $\sim10$ Myr and difficult to find in the field.  However, in many cases these necessarily wide pairs, as judged by some of the age criteria discussed here, were optical doubles, not physical binaries \citep{palla92, martin92}, although different studies do not agree on which systems may be physical \citep{gerbaldi01}.

A variation on binaries as mini-clusters occurs when one of the stars is a white dwarf.  \citet{catalan09} have used WDs in wide binary systems as a check on ages of low-mass stars.  Wide binaries (with separations of $\sim1000$ AU or more) are used as test particles for age-related dynamical processes, both in star-forming regions (e.g., \citet{kraus09a}) and among the older populations of the disk (e.g., \citet{makarov08}).  For those systems one can help weed out non-physical pairs with independent age-related criteria.

Other special uses of binaries are discussed in the next section on OC ages: eclipsing binaries as mass calibrators and tidal effects as an age-ordering method.

In some instances binary systems may be misleading.  As just noted, one must first apply astrometric and radial velocity criteria to ensure the two stars form a physical pair.  But even when they do the masses we see now are not necessarily the same as when the system formed.  For example, \citet{rapp09} show that the primary in the Regulus system has a significantly different mass now than it did in the past; Regulus is one of the Lindroos systems just noted and this fact, of course, substantially alters conclusions one might draw about the companion K dwarf.  In another potential example, \citet{stassun08} note that the two components of the PMS system Parenago 1802, in the Orion Nebula Cluster, have the same mass yet substantially different temperatures and luminosities.  They suggest that the two stars may have formed at different times and that is possible, particularly if companions may have been exchanged as two binaries encountered each other.  However, this system has an orbital period just under 5 days which would make it resistant to dynamical effects but, at the same time, the two components of this system in their earliest stages were significantly larger than they are now and their propinquity may have led to effects that produced the current differences in \teff\ and \Lbol.  Similarly, the brown dwarf eclipsing binary found by \citet{stassun07} may be in a different state now than when it formed; its orbital period is just under 10 days.  Things are not always what they seem.


\section{OPEN CLUSTER AGES}\label{OC-age}

OCs provide the benchmarks to calibrate virtually every stellar age except that of the Sun.  The essential tool for studying clusters is the HRD --  \loglbol\ versus \logteff\ -- and its observational counterpart, the CMD.  HRDs are powerful tests of stellar physics because the models have to work consistently over a very broad range of mass and evolutionary states, where many different physical processes are taking place.  At the same time, that means that derived ages to some extent themselves depend on the cluster age.  For instance, very young clusters usually have ages determined from their massive stars at the MSTO, whereas intermediate-age clusters have ages determined from intermediate-mass stars where the interior physics is different.  Old clusters' ages often depend on the locations of evolved stars in the HRD.

For the present purposes, the relevant OCs are the nearby well-studied examples (within about 1 kpc).  The more distant clusters are important for understanding our Galaxy, particularly because old OCs are inherently rare, but it is primarily the nearby OCs that will be discussed here because their low-mass members are bright enough to be accessible to spectroscopy and with that a fuller understanding of their properties.  In most cases there is ample observational information to assess the membership of individual objects well, reducing systematic effects from field star contamination, and the metallicity of the cluster is known from spectrum analyses.  These are some of the problems and limitations that are considered in analyzing the ages of the nearby OCs:

\begin{enumerate}

\item Reddening and extinction are always uncertain, but for the nearby OCs those quantities are small and so the errors tend to matter less.  However, PMS clusters and associations are generally still located in and amid the material from which they formed and their reddenings can be high and significantly different from star to star.  The reddening law to apply also may vary from place to place and is a source of systematic uncertainty.

\item Both \Lbol\ and \teff\ are theoretical quantities that are useful for describing models, but are simplifications when considering real stars.  In particular, \teff\ is a parameter that describes the energy flux of a star through its surface, based on calculations of the energy generation occurring in the stellar interior, but \teff\ as a single number does not describe the way that flux appears in observations.  Stellar surfaces are complex and the formation of spectrum lines and spectral energy distributions are also complicated when considered in detail.  A star like the Sun, for example, would need three components to describe its surface even in a fairly simplistic way: the upwelling hot material, the cool material moving downward, and the spots, each with an areal coverage and a temperature or temperature gradient.  Chromospheric activity also changes the temperature structure in the outer parts of a star's atmosphere, altering the line formation conditions.  It is this inherent complexity that likely accounts for how different photometric indices predict different temperatures, none of which agree with what comes from spectrum analysis.  Comparisons of different color-temperature relations (CTRs) with each other and with \teff\ values determined from spectrum analysis show differences of 80--100 K \citep{ramirez05} and 100 K is generally taken to represent the true uncertainty of stellar \teff\ values.  The study of \citet{ribas09} is illustrative in that they derive \teff\ values for a single, well-observed star using a variety of methods (photometry, ionization balance, excitation balance) without finding a single, consistent value.  A part of the problem is that different color indices are sensitive to different aspects of a star's atmosphere.  In particular, \bv\ appears to be sensitive to spots on young G-K dwarfs \citep{stauffer03} and the \vi\ index appears to produce more reliable \teff\ values (\citet{stauffer03}; \citet{lyra06}).

\item OC ages derived from different methods with different underlying physical processes can yield ages that different systematically to a significant degree.  A notable example are ages from the lithium depletion boundary (LDB) in very young clusters and groups (see below).  Five LDB ages are now available and all are systematically about 50\% larger than ages determined from the MSTO.  The age scale from the LDB involves fundamentally simple physics, whereas MSTO ages rely on knowledge of the interiors of massive stars where processes may be taking place that we do not fully appreciate.  Because of this, the LDB age scale may be the more reliable, but there are too few LDB ages to recalibrate the overall OC age scale (see Sec. \ref{LDB}).

\item The transformation of magnitudes to \loglbol\ requires knowledge of the cluster's distance, the extinction, if any, and bolometric corrections.  Only a very few OCs have trigonometrically determined distances, and even some of those remain contentious.  Indeed, the process of fitting isochrones to OC CMDs generally yields a best-estimate distance as part of the exercise.

\item Because the initial mass function for clusters is strongly biased against high-mass stars, there are generally very few stars in young clusters to define the MSTO and this is exacerbated by the presence of binaries.   \citet{jorgensen05}, \citet{naylor06} and \citet{mayne08} have attempted to reduce systematic errors in MSTO fitting by applying bayesian techniques.  Similar efforts for many more of the nearby clusters could at least lead to a more consistent age ordering.

\end{enumerate}

\citet{mermilliod00} reviewed age estimation methods for OCs and included some of the techniques that use HRD morphology that are mainly applied to old clusters.  \citet{gallart05} and \citet{meynet09} provide a thorough discussion of the problems faced in determining cluster ages from isochrones, particularly a close examination of stellar models.  That review is oriented toward disentangling the star formation histories of nearby galaxies from their composite CMDs, but the limitations and problems that arise in applying the models are discussed extensively.  For example, an age derived from a MSTO can change by up to 50\% depending on the specific models used, but that difference is greatest in certain age ranges and very low in others.  This is a reflection of the different non-standard physical processes that may be included by different groups because those processes take place within specific mass ranges, which is to say age ranges, given the obvious connection between the masses of stars at the MSTO and the cluster age.  For some particular well-studied cases, \citet{pinsono04} and \citet{an07} have been able to construct empirical isochrones for ZAMS and post-ZAMS OCs, and \citet{mayne07} have done so for PMS stars.  This was done to lessen problems with the CTRs and led to improved distance estimates but did not directly affect derived ages as much.

The physical processes that are not well understood include convective core overshoot, rotationally-induced mixing, internal gravity waves, and diffusion.  Magnetic fields may also play a role.  This review cannot treat this subject in any detail, but what is important here is that several of these processes -- convective core overshoot and rotationally-induced mixing in particular -- replenish fuel in the nuclear-burning regions of stars and so can significantly extend the MS lifetimes of massive stars by up to $\sim50\%$ \citep{rosvick98, woo01}.  This leads to fundamental and significant uncertainty in the overall age scale of clusters, especially young clusters, and of stars in general.

Very young clusters present their own challenges.   For any method, the uncertainties in both \Lbol\ and \teff\ are significantly larger for PMS stars and this has many causes, some observational and some inherent to the stars themselves; see Sec. \ref{PMS}.   Along the upper MS, \citet{mayne07} discuss a ``radiative-convective'' gap where stars develop their radiative core as they first reach the ZAMS; this gap is very close to where the PMS line for low-mass stars meets the ZAMS.  The R-C gap moves in the CMD with age and allowed \citet{mayne07} to set an age ordering for very young clusters that is different from the age order determined from the MSTO.  However, the R-C gap is a low-contrast feature in the CMD that can be difficult to perceive in sparse clusters.  \citet{mayne08} applied objective criteria \citep{naylor06} to MS fitting of very young clusters to avoid the biases that come with fitting by eye.  A major goal of these studies is to determine the extent of age spreads within star-forming regions, and that will be discussed in the next section as well.

PMS clusters can sometimes have conflicting age indicators, as when massive stars are leaving the MS at the MSTO and low-mass stars are just reaching the ZAMS.  \citet{lyra06} have found that such discrepancies depend on the color index used, as noted above, and that the MSTO and MS turn-on ages are in good agreement when colors such as \vi\ are used that avoid the blue part of the spectrum.

Old clusters also pose difficulties.  Even for well-studied cases such as M67 and NGC 188 there can be field star contamination among the faint stars of the lower main sequence and the degree of convective core overshooting remains uncertain \citep{salaris04, VDB04}.

\subsection{The Lithium Depletion Boundary}\label{LDB}

The potential significance of these alterations of the OC age scale is illustrated by the ages derived from the location of the lithium depletion boundary (LDB) seen in some young clusters.  The LDB was first proposed as a means of discriminating bona fide brown dwarfs (BDs) from very-low-mass stars by \citet{rebolo92}.  They pointed out that objects below about 0.06 \MS\ will never get hot enough in their cores to destroy their Li, which happens at temperatures above about 2.5 MK.  The temperature and luminosity of that dividing line changes with age as the 0.06 \MS\ stars cool and as less massive objects acquire the core temperature needed to astrate Li.

\citet{basri96} were the first to detect the LDB in a cluster.  They detected Li in a single very faint Pleiades object and applied the models of \citet{nelson93} to derive a cluster age of 110-125 Myr, substantially greater than the 70 Myr that results from fitting the MSTO  \citep{stauffer98}.  Subsequent work has refined the models \citep{burke04} and has added new observations that bring the number of detections of the LDB to five clusters: the Pleiades \citep{stauffer98}; $\alpha$ Persei \citep{stauffer99}; NGC 2547 and IC 2391 \citep{jeffries05}; and IC 4665 \citep{manzi08}.

For the youngest clusters the absolute differences in age between MSTO and LDB determinations are small, yet the LDB ages in all cases are systematically about 50\% greater than MSTO ages.  The observations needed to detect the LDB in a cluster require good spectra of extremely faint objects, limiting the test to the nearer OCs.  Yet the LDB ages pose a real problem, and that is because the physics in the models that predict the position of the LDB with age depends only on the core temperature reached and that is calculated straightforwardly.  The surfaces of BDs may be more complex than we can understand at this time, but that doesn't influence what occurs in the object's core.  For these reasons the age scale implied by the LDB observations may be more reliable than that from the MSTO, but there remain too few clusters with LDB ages to readjust the scale of MSTO ages, and that situation is not likely to change for some time.  

It is also possible, however, that our estimates of the radii of very young stars are systematically wrong, and this would affect the interpretation of LDB ages.  There is accumulating evidence that PMS and ZAMS stars are larger than models predict (\citet{berger06}; \citet{lopez07}; \citet{ribas08}; \citet{torres09}; \citet{jackson09}) and this makes them appear cooler; adjusting for this effect may help to bring LDB and MSTO ages of OCs into agreement \citep{yee10}.  Other evidence suggests the MSTO ages are sound.  For instance, \citet{makarov06} was able to precisely fit a model to the star $\alpha$ Per itself that is consistent with a good fit to the MSTO in the $\alpha$ Persei cluster for an age of 52 Myr, as compared to an LDB age of $90\pm10$ Myr \citep{stauffer99}.

\subsection{Other Indicators of Cluster Ages}

\subsubsection{TIDAL EFFECTS ON BINARY ORBITS}

Tidal interactions between the stars in close binary systems influence the evolution of the stars themselves if they are close enough but more generally lead to changes in the orbit, including synchronization of the stellar rotation with the orbit and circularization of the orbits.  This is especially effective in low-mass stars because they can dissipate angular momentum in their winds and thus remove AM from their orbits, leading to coalescence.  The theory of these interactions is not important here; what matters is that we expect to see a progression with cluster age of the period at which synchronization and circularization are seen and that is largely borne out observationally \citep{meibom05, meibom06}, although the Hyades and Praesepe appear to conflict with the trend.  

Whether or not the physics of tidal dissipation is fully understood, the process can only act in one direction and so observations of these tidal effects should make it possible to place OCs in a reliable age ordering.  In practice the observations required are intensive (precise photometry and spectroscopy over extended times to get rotation periods and orbital parameters) and limited to nearer clusters.

\subsubsection{WHITE DWARFS IN CLUSTERS\label{WDs}}

The detection of the white dwarf cooling sequence in a cluster allows for a determination of the cluster's age.  This subject has been well reviewed elsewhere (\citet{koester02}, \citet{vonhippel05}, \citet{winget08}, \citet{salaris09}) and so only a few comments will be made here.  

First, the physics of WD cooling itself seems straightforward because no energy is being generated; there is just gradual cooling of the star with time and the physics of that process is fairly well understood.  However, there are uncertainties that matter for the input physics, such as in the equation of state, opacities, etc.  The detailed energy budget is strongly dependent on the chemical stratification of both the degenerate core and the envelope.  All of these effects (and more) mean that some claimed uncertainties for cluster ages from WD cooling sequences are probably optimistic, and there is an inherent error of $\sim1$ Gyr.

Second, to derive the cluster age also requires determining the initial-final mass relation, which relates the WD mass to the mass of the progenitor star prior to mass loss on the asymptotic giant branch.  There are inherent uncertainties in this determination, such as the degree of mass loss on the asymptotic giant branch.

Third, the above uncertainties arise from the models, but there are additional observational factors (including composition and temperature); see \citet{salaris09}.

Fourth, the usual technique is to detect the coolest WDs in a cluster but that can be difficult and can lead to selection effects.  \citet{jeffery07} discuss a means of using brighter WDs in clusters to the same end.

Finally, the cluster as seen today can be the result of dynamical evolution that influences the conclusions drawn.  An example is the WD age determined for the Hyades \citep{weidemann92}, which, at 300 Myr, is about half the MSTO age of 625 Myr \citep{perryman98}.  This discrepancy has been argued to be due to the coolest WDs having been ejected from the cluster.

\subsubsection{ECLIPSING BINARIES}

Eclipsing binaries (EBs) play a special role in studying stellar astrophysics and OCs \citep{stassun09}.  If the components are well separated so that mass exchange or other physical interactions are unlikely to have taken place, then finding EBs in OCs and measuring their orbits allows one to place a known mass benchmark on the models.  In addition, EB orbits yield radii, another fundamental property of the models.  As a result EBs can provide critical tests of model isochrones \citep{southworth06}, particularly in the fortuitous case of an EB being found near the MSTO.  Also, EBs allow accurate distances to be determined, even for nearby galaxies \citep{guinan98, clausen04}, and those distances are reddening- and extinction-independent.

For instance, \citet{grundahl08} studied an EB in the metal-rich old cluster NGC 6791, and in NGC 188 \citet{meibom09} have analyzed an EB near the MSTO.  In both cases it was possible to find a precise age (to about 2-3\%) for the EB by comparing to theoretical isochrones in a mass-radius diagram, independent of distance, reddening, and a color-temperature relation.  This enabled them to critically test conventional isochrones and their sensitivity to metallicity, distance, and reddening.  The ages that result from analyzing EBs are still model-dependent, but offer an additional means of testing stellar physics.  In a similar fashion, \citet{clausen09} were able to use models to get an age for the field EB V636 Centauri, a pair of solar-type stars.  This was especially important because the radii for this system were at odds with conventional models, which, along with other evidence, has suggested that young, active solar-type stars have larger radii than has generally been assumed (see comments in Sec. \ref{LDB}).

\subsection{OC Age Precision and Accuracy}

For reasons noted above there remains significant uncertainty in the accuracies of OC ages.  Even ignoring discrepancies between different methods, the standard technique -- isochrone fitting -- has a precision of $\sim10$\% \citep{meynet09}.  As an example, the Hyades presents what has to be the most favorable case possible, with zero reddening, a well-determined metallicity, only a very small uncertainty in distance, a fairly rich population, and unambiguous membership and multiplicity.  Even with these significant advantages the best age is $625\pm50$ Myr \citep{perryman98}.  As another example, \citet{jorgensen05} applied a bayesian fitting method to the MSTO in two clusters: IC 4651 (deriving $\tau = 1.56\pm0.03$ Gyr) and M67 ($\tau = 4.05\pm0.05$ Gyr).  These  are remarkably small errors ($\sim1-2$\%) but do not take into account the full range of possible uncertainties.  The uncertainties usually quoted for OC ages are $\sim10$\% and that probably represents a good estimate of just the fitting errors. 


\section{A COMPARISON OF TECHNIQUES AND A SUMMARY}

How well do results from different methods compare?  First, we note that some methods are inherently limited to being applied to ensembles of stars, while others are for individual stars.  For individual stars, the available methods include isochrone placement; nucleocosmochronometry; asteroseismology.  The methods limited to ensembles include:

\begin{itemize}
\item The lithium depletion boundary (LDB) method for young clusters, because it depends on finding the point among a group of low-mass objects where lithium reappears.
\item Kinematic expansion, because it involves determining when in the past a group of stars occupied the least space.
\item The empirical methods -- rotation, activity, and lithium -- when applied to young stars ($\tau < 0.5$ Gyr) because the inherent scatter among coeval populations is large and exceeds the change in the quantity over significant times.
\item Isochrone fitting to OCs, for obvious reasons.
\end{itemize}

And the empirical methods (declines in rotation, activity, and Li) can be used for either individual stars that are $>0.5$ Gyr or ensembles that are younger.

\begin{figure}[t]
\includegraphics[scale=0.90]{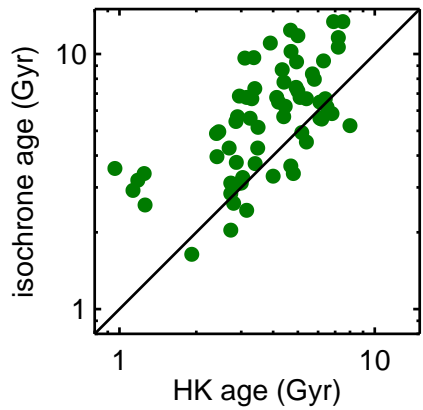}
\caption{\label{HK-isochrone}
A comparison of ages determined from an HK activity index and from isochrone placement.  This sample consists of $\sim100$ solar-type stars that have detected planetary companions.  The HK ages are from \citet{mamahill08} and the isochrone ages from \citet{takeda07}.  Note that the isochrone ages, on average, are about 1.5 times the HK ages.
}
\end{figure}

There are not many instances of stars with ages determined in multiple ways so that one can compare results from the various methods.  Also, methods such as using rotation and activity are not really independent of one another and show an inherent physical correlation.  However, Figure \ref{HK-isochrone}
\marginpar{\bf Fig. \ref{HK-isochrone}: HK ages versus ages from isochrone placement.}
compares ages of G dwarfs derived in two very different ways: from isochrone placement (Sec. \ref{isochrone-MS}) and from rotation rates (Sec. \ref{spindown}).  The lack of a perceptible correlation between the two methods is not encouraging.  \citet{lachaume99} derived ages for a number of nearby main sequence stars using available information (isochrone placement, rotation, HK emission, space motions, and metallicities) and similarly saw poor agreement among the indicators.


\section{THE FUTURE OF AGE}\label{future}

If this review has seemed downbeat on the prospects of determining ages of individual stars that is because there is a lot to be negative about and I have wanted to ensure that readers appreciate how difficult the task is.  Few topical symposia have been held on the subject of ages; the first was ``l'Age des Etoiles,'' held in Paris in 1972.  A look through the proceedings of that IAU colloquium \citep{cayrel72} shows that many of the subjects discussed there have advanced rather little.  We now have much more and much better observations, but many of the same problems remain.

Also, one desires to determine an age with a precision comparable to other stellar quantities and there are inherent factors that prevent that.  Nevertheless, age-dating is not an impossible task, just a slippery and uncertain one, and it is possible to establish at least limits on how old most objects are.  Despite what I have said, I am very optimistic about the future.  How can we improve on matters, and what does the future hold?  What can be done to improve the situation?

First, the near future is very promising, in large part because of the {\em CoRoT} and {\em Kepler} missions.  In particular, {\em Kepler} is obtaining ultra-high-precision (1:10$^5$) photometry for $\sim100,000$ solar-type stars in order to detect the transits of Earth-sized planets.  In addition to what it does for planet-hunting, {\em Kepler} will leave a rich legacy for stellar astrophysics, for never before have we been able to observe stars at such a precise level (excepting the Sun itself, of course, and the advent of similar quality photometry for the Sun is very recent).  The {\it Kepler} data will be like ``full body scans'' for solar-type stars and should enable the detection of not just rotation in older stars but also differential rotation as the apparent \prot\ changes.

Of greatest relevance here is the fact that {\em Kepler} obtains its photometry at one-minute intervals for several thousand selected, brighter stars in its field (the remainder are sampled in 30-minute integrations).  High photometric precision combined with long on-target time sequences will enable the detection of solar-like oscillation modes in these stars.  Those asteroseismological detections will enable reasonably precise ages to be assigned, and most of those stars will be older stars of the field.  Those ages will only be as good as the models used to calculate stellar properties, but, at the same time, the physical constraints imposed by asteroseismology will stress those models and help us refine them.  In addition, {\em Kepler}'s high precision should make it possible to detect, for instance, those stars' rotation periods, even for the old, inactive stars.  The apparent \prot\ changes as spots migrate, as they do on the Sun, and so good photometry can also enable us to measure stellar differential rotation with precision, a critical parameter for dynamo models.  With sufficient devoted time on selected targets (months), asteroseismology can also show the signature of the convection zone depth, the first time we will have such a key constraint on models for any star other than the Sun.  {\it Kepler} is a solar physics mission at heart.  These ages from asteroseismology for older stars will be a breakthrough in that they will enable much better testing of the empirical methods and should help us understand if ages from isochrone placement and the empirical methods are sound.

Another key mission in the near future will be {\em Gaia}, which will obtain highly precise ($\sigma \approx 100 \mu$arcsec) parallaxes for millions of stars in our Galaxy.  {\em Gaia} should remove almost all ambiguity about the true distances to any of the open clusters used as calibrators and will also help weed out non-members in those clusters.  With results from {\em Gaia}, luminosities for the nearer stars will be virtually error-free except for uncertainty in bolometric corrections.  {\it Gaia} will also provide accurate distances to PMS objects in star-forming regions, again removing ambiguity and helping us to understand those stages of stellar evolution.

The proposed variability surveys, such as the {\it Large Synoptic Survey Telescope}, will provide a bounty for studies of stellar variability.  More important, many eclipsing binaries will be discovered, enough so that we can choose those best suited to test stellar models.  In general these missions improve our knowledge and understanding of our Galaxy by probing the data domain along dimensions not yet explored.

Our ability to study, understand, and model the physics of stars has improved enormously over my career.  The new domains of astronomical observation that are becoming available will only challenge what we think we already know and will lead to substantial improvements.  Each new domain we explore reveals the weaknesses of the models and forces improvements. 

\subsection{Predictions}

If I were any good at predicting you would have heard about it before now.

\begin{enumerate}

\item {\it Kepler} will produce excellent asteroseismological data for a large number of solar-type stars and that, combined with additional observations and careful modeling, will yield better ages for older stars than we have ever had.  Those ages will enable us to calibrate and understand empirical age indicators such as rotation and activity.  Rotation will turn out to be as well-behaved as we have assumed.  Activity will be useful statistically for stars more than $\sim1$ Gyr old, and we will gain some knowledge of the inherent variation and scatter in activity at different ages.  Lithium will remain confusing and resistant to modeling.  {\it Kepler} will also for the first time turn asteroseismology from a boutique operation into a wholesale one, with good frequencies available for thousands of stars.  

\item Critical physical information, such as the depth of stellar convective envelopes, will result from long on-target integrations with {\it Kepler}.  When combined with concurrent observations of chromospheric activity we will gain insight into the factors that drive dynamos and long-term activity cycles.  If we can understand the physics of the solar 11-year cycle we can start to understand stellar dynamos in general and we can turn rotation and activity from empirical age indicators into model-dependent ones.

\item Suitable binaries will be found in young clusters enabling accurate masses to be measured for ZAMS solar-type stars.  Those masses will be significantly different than the models have predicted for a given location in the HRD and will lead to important insights in stellar models.  This may hold the key to understanding the scatter seen in rotation, activity, and lithium in young clusters, as well as the discrepancy between ages from the main sequence turn-off and the lithium depletion boundary.  The large scatter seen for rotation, activity, and Li in PMS and ZAMS clusters will be explained as being due to the distribution of angular momentum, disks, and companions when stars are formed and the later consequences of those.

\item The best-available data for the nearest OCs will be analyzed and fitted in a self-consistent manner to produce at least a more reliable age ordering for those clusters.  This will lead to some insights on what factors produce differences between clusters that we do not now understand, such as conflicts between MSTO ages and an age determined from PMS isochrones.

\item Knowledge of age and mass will enable us to create a sample of bona fide solar analogs and we will see that the Sun is a very typical star.  Accurate ages will make it possible to find a solar-age star in a ``Maunder minimum'' state and that will help us understand better the connection between changes in luminosity and activity for stars like the Sun.

\item An Earth-like planet will be found around a nearby solar-type star and that planet will show a biomarker.  We will estimate the star's age from asteroseismology and it will be $>2$ Gyr old.

\end{enumerate}


\section{Acknowledgements}

The ages of solar-type stars have been in my mind since my thesis.  Those who have motivated me in this have themselves been instrumental in examining the Sun among the stars, the life history of the Sun, and what the cosmos tells us about the formation and evolution of our solar system.  I especially want to acknowledge George Herbig, my thesis advisor, for exceptional patience and a broad view of astronomy; and Jack Eddy, for showing that we can learn about the Sun from information right here on Earth, on the Moon, and beyond.  Olin Wilson deserves recognition as well for his long-term study of activity cycles on stars and Giusa and Roger Cayrel have also been pioneers in the detailed study of late-type stars.

This review had its start in Symposium 258 of the International Astronomical Union, ``The Ages of Stars,'' held in Baltimore, Maryland, in October, 2008.  I had wanted to hold such a meeting for a long time because the subject of stellar ages is so interesting and yet so little understood.  That symposium was, in part, made possible by support from the IAU, the U.S. National Science Foundation, the Las Cumbres Observatory Global Telescope Network, and the Space Telescope Science Institute.  The meeting brought together about 150 astronomers from many countries to address the subject of ages in a broad sense, over most stellar types, and including populations as well.  All those who came and especially those who helped organize the symposium have my gratitude for turning the symposium from an idea into reality and for teaching me so much.  I look forward to attending the next similar symposium to see what progress has been made.

In more recent times I have profited from interacting with many people, too many to list fully, but I wish to particularly thank Jeff Valenti, Corinne Charbonnel, and Georges Meynet.  I also note with gratitude the help of J. Irwin, J. Hall, J. Valenti, N. Da Rio, M. Robberto, E. Mamajek, and M. Slipski in providing data for the illustrations.  The careful reading of S. Faber improved the presentation significantly.

Parts of this review were written during an extended stay at the Observatoire de Gen\`{e}ve.  I thank Corinne Charbonnel for the invitation and Gilbert Burki, the Director, for the support provided.  A research leave from STScI is also gratefully acknowledged.



\section{Sidebar 1: Why Care About Stellar Ages?}\label{sidebar1}

Some day soon we anticipate the detection of Earth-sized planets around stars.  Before too long, we hope, we can expect that someone will identify a sign of life -- a biomarker -- on an Earth-like planet around another star.  Given the need for high angular resolution to make that observation and the inherent paucity of photons to work with, the star in question will be one close to us, which is to say a field star that is unassociated with a cluster.  When that claim is announced, the first question we will all ask is "How old is that star?",  so that we can assess the planet's evolution.  Yet determining that age is likely to be difficult and imprecise, as problematic in its own way as the observation of the biomarker.

\section{Sidebar 2: A Challenge for Stellar Astrophysics}\label{sidebar2}

For many years there has been tension between modelers of stellar structure and cosmologists in that the modelers tended to derive ages for the Galaxy's oldest clusters that approached or exceeded the apparent age of the Universe.  For much of that time the stellar astrophysicists had the advantage of better-constrained physics, but the recent results from the WMAP mission lead to an age for the Universe of $13.7\pm0.2$ Gyr \citep{bennett03}.  In other words, cosmologists can now claim to determine the age of the Universe to within about 1\%.  By contrast, the ages of individual nearby stars can be estimated to no better than 10\%, and that uncertainty is probably optimistic and does not fully take into account systematic effects.  Stellar astrophysicists should see the WMAP results as a friendly challenge that we can meet and exceed, but only after significant exertion.  We should be able to understand the closest stars at least as well as we can the entire cosmos!

\section{Sidebar 3: The Sun as a Benchmark}\label{sidebar3}

There is exactly one stellar age that is both precise and accurate, that of the Sun, and it illustrates some of the inherent problems in determining ages.  The Sun is $4,567\pm1\pm5$ Myr old \citep{chaussidon07}.  The extraordinary precision of 1 Myr represents measurement error (individual measurements are precise to 0.6 Myr \citep{amelin02}), and the only slightly larger systematic error of 5 Myr is due to uncertainty over the precise sequence of events in the early years of the solar system's history.  That systematic error should lessen as we understand those events better.  This age is determined from the decay of radionuclides.

The Sun itself does not reveal its age in any of its observable properties.  It is only because we can measure solar system material in the laboratory that we can establish the Sun's age with complete confidence; that is not possible for any other star.  The Sun thus forms the only age benchmark available in all of stellar astrophysics, with all others being model-dependent to some extent.  The extraordinarily precise age of the Universe from WMAP ($13.7\pm0.2$ Gyr; \citet{bennett03}) is based on some assumptions, but the accuracy is likely better than 10\% at the very least.

The Sun matters in another crucial respect, and that is for its absolute abundance scale.  Abundances in stars are measured relative to the Sun, but the abundances that determine the structure of stellar models are absolute values.  The recent controversy over CNO abundances in the Sun matters for ages because changing solar absolute abundances requires changes in all the models.

For more on the Sun as a fundamental calibrator, see \citet{JCD09}.


\section{LITERATURE CITED}

\section{FIGURE CAPTIONS}

\end{document}